\documentclass[fleqn,usenatbib]{mnras}
\usepackage[utf8x]{inputenc}
\usepackage[T1]{fontenc}
\usepackage{ae,aecompl}
\usepackage[english]{babel}
\usepackage{lmodern}
\usepackage{enumitem}
\usepackage[squaren, Gray, cdot]{SIunits}
\usepackage{multirow}
\usepackage{color}
\usepackage{array}
\usepackage{setspace}
\usepackage{graphicx}
\usepackage{amsmath}
\usepackage{amssymb}
\usepackage{bm}
\usepackage[super]{nth}

\definecolor{Blue}{rgb}{0,0.08,0.65}
\definecolor{Red}{rgb}{0.65,0.08,0.05}
\definecolor{Green}{rgb}{0.15,0.45,0.25}
\definecolor{Purple}{RGB}{153, 51, 153}
\def\red#1{#1}

\title[Post-Born convergence PDF]{Post-Born corrections to the one-point statistics of (CMB) lensing convergence obtained via large deviation theory}

\author[A. Barthelemy et al.]{
Alexandre Barthelemy$^{1}$\thanks{E-mail: alexandre.barthelemy@iap.fr},
Sandrine Codis$^{1}$, Francis Bernardeau$^{1,2}$
\\ 
$^{1}$CNRS \& Sorbonne Universit\'e, UMR 7095, Institut d'Astrophysique de Paris, 75014, Paris, France\\
$^{2}$Institut de Physique Th\'eorique, Universit\'e Paris-Saclay,
CEA, CNRS, UMR 3681, 91191 Gif-sur-Yvette, France
}

\date{Accepted XXX. Received YYY; in original form ZZZ}

\pubyear{2020}

\begin{document}

\label{firstpage}
\pagerange{\pageref{firstpage}--\pageref{lastpage}}
\maketitle

\begin{abstract}
Weak lensing of galaxies and CMB photons through the large-scale structure of the Universe is one of the most promising cosmological probes with upcoming experiments dedicated to its measurements such as Euclid/LSST and 
CMB Stage 4 experiments. With increasingly precise measurements, there is a dire need for accurate theoretical predictions. In this work, we focus on higher order statistics of the weak lensing convergence field, namely its cumulants such as skewness and kurtosis and its one-point probability distribution (PDF), and we quantify using perturbation theory the corrections coming from post-Born effects, meaning beyond the straight-line and independent lenses approximations. At first order, two such corrections arise: lens-lens couplings and geodesic deviation. Though the corrections are small for low source redshifts (below a few percents) and therefore for galaxy lensing, they become important at higher redshifts, notably in the context of CMB lensing, where the non-gaussianities computed from tree-order perturbation theory are found to be of the same order as the signal itself. We include these post-Born corrections on the skewness into a prediction for the one-point convergence PDF obtained with large deviation theory and successfully test these results against numerical simulations. The modelled PDF is indeed shown to perform better than the percent for apertures above $\sim 10$ arcminutes and typically in the three sigmas region around the mean.
\end{abstract}


\section{Introduction}

When trying to extract (cosmological) information from cosmic fields, it is common to focus on power spectra (in Fourier space) or two-point correlation function (in real space) but those observables contain only complete statistical information for pure Gaussian random fields. However, even starting from Gaussian initial conditions, the subsequent non-linear time evolution of matter density fluctuations develop significant non-Gaussianities, in particular for small scales and late times. This justifies the need for the theoretical development of non-Gaussian observables, in particular with the advent of future large galaxy surveys. Many different strategies have been put forward over the course of the last decades for which we mention recent applications, such as peak counts \citep{KacprzakDES16,2018MNRAS.474.1116S}, Minkowski functionals \citep{2019JCAP...06..019M}, voids \citep{voids} or one-point probability distribution function (hereafter PDF) \citep{GruenDES17}. In particular, PDFs are shown to be a very interesting probe that, on the one hand, can be fully modelled from first principles \citep{saddle} and, one the other hand, can be easily applied to real data offering potentially much more information \citep[see for instance][for an illustration on the dark energy equation of state]{encircling} that can help break some degeneracies as was shown for instance in \cite{Uhlemann19} notably for the case of the mass of neutrinos. 

Given the promises of cosmic field PDFs, recently the one-point PDF of the lensing convergence in cones in the quasi-linear regime was modelled from first principles using large deviation theory in \cite{Barthelemy19} -- which is strictly speaking equivalent to computing all the cumulants of the field at tree-order in perturbation theory as was shown in \cite{Bernardeau1992} and \cite{Valageas}. 
However in \cite{Barthelemy19} the convergence was modelled as a weighted integral over the underlying non-linear density field -- therefore assuming independent lenses -- along the un-perturbed line-of-sight. Hence, two ingredients were neglected: i) couplings between the lenses which state that the combination of lenses in geometrical optics is not linear, and ii) the fact that background lenses are themselves lensed by foreground lenses, thus changing the overall trajectory of light rays. 
These terms will tend to Gaussianise the convergence field since they characterise the introduction of random deflections along the light path which will in turn tend to diminish the impact of the non-linear clustering of matter. The mind picture one could form is that of clustered chunks of matter blurred by this lensing terms.

Those corrections, often called post-Born corrections in the literature, have been studied for a long time, we refer the reader to classical weak-lensing reviews and textbooks for more details (see for example \cite{ReviewBartelmann}). Note that at second order in the gravitational potential, \cite{Bernardeau10} computed all corrections to the shear field in a full sky relativistic setup and found that all other terms apart from post-Born corrections are strictly general relativistic and only important at large angular scales which does not preoccupy us here. We will therefore focus only on post-Born and non-linear corrections here.

At the level of cosmic shear experiments, that is for redshift of source galaxies of order 1 and sufficiently large scales, 
the power spectrum of the convergence field is very insensitive to post-Born corrections, which, as shown by \cite{Shapiro06}, are several orders of magnitudes below the signal, and only become more important than cosmic variance for a full-sky survey at really small scales down to which other uncertainties arise such as baryonic physics. This was confirmed for instance by \cite{Hilbert09} with the help of ray-tracing through N-Body simulations which showed that the first-order (in the Newtonian potential) approximation provides us with an excellent fit to cosmic-shear power spectra as long as the exact (fully non-linear) matter power spectrum is used as an input. Using the same simulations, cosmic-shear B-modes,  which  are  induced  by  post-Born  corrections  and lens-lens coupling, were also found to be at least three orders of magnitude smaller than cosmic-shear E-modes. Going beyond the convergence and shear, \cite{Schafer12} focused on the cosmic flexion and showed that post-Born corrections to the flexion power spectra are roughly four order of magnitudes below the signal at redshift of order 1, again much below the cosmic-variance limit. As for non-Gaussian statistics, \cite{Bernardeau1997} for the convergence and \cite{Schneider1998} for the aperture mass both showed that the correction induced on the skewness was also negligible for sources at redshift around 1, the corrections being smaller than the errors induced by using the tree-order perturbation theory result to get the skewness, which was also confirmed extracting those skewness corrections from numerical simulations \citep{Petri17}. Hence making use of the standard definition of weak-lensing quantities neglecting post-Born corrections is supposedly enough for the majority of weak-lensing experiments as long as very small scales are not considered. Also note that the addition of baryons does not change this picture \citep{Celine}.

In the context of CMB lensing, post-Born corrections also have a negligible effect onto power spectra: they appear to be irrelevant for the E-mode temperature and polarisation power spectra and differences on the B-modes power spectra only show up at very small scales ($\ell \geq 3000$) \citep{Pratten16,Marozzi18,Fabbian18}. 
Similar findings are reported on the cross-correlation between the thermal Sunyaev-Zeldovich effect and the observed convergence (with supplementary terms coming from the reduced shear), the correction being several order of magnitude below the first-order result and under the cosmic-variance limit up until very large angular scales ($\ell \sim 4000$) as shown by \cite{Troster14}. However, post-Born effects might be important for other CMB lensing observables, for instance when cross-correlating galaxy counts and CMB lensing convergence \citep{Bohm19,Fabbian19} or when cosmic shear maps is used to reconstruct the lensing potential power spectra for which a significant bias appears when neglecting both post-Born terms and the non-Gaussianity of the large-scale structure \citep{Bohm16, Bohm18, Beck18}. Post-born corrections also seem important for higher-order statistics of CMB lensing. As an example, their contribution to the CMB convergence bispectrum is at a comparable amplitude than the signal coming from gravitational non-linearities \citep{Pratten16,Fabbian18} and therefore has to be accounted for if bispectrum is used to constrain possible modifications of gravity \citep{Namikawa18} although this might not be observable even with stage-IV type experiments  \citep{Namikawa19}.

In this work we focus on the convergence one-point PDF which was showed to be of significant importance by \cite{JiaCMB} since the covariance between PDF and power-spectrum is relatively small so that there is information to be gained by studying the PDF. Its study notably provides an improvement of 30\% on the $\Omega_m-\sigma_8$ error contour once added to the power spectrum. We specifically focus on adding post-Born corrections from first principles into the large-deviation formalism derived by \cite{Barthelemy19}. To this aim, we first re-derive in Section~\ref{sec:postborn} the optical Sachs equation from basic general relativity, and the linear deformation matrix when solving the equation on a linearly perturbed flat Friedman-Lemaitre-Robertson-Walker (FLRW) spacetime. 
We then introduce Newtonian non-linearities in the matter density field and coupling terms between lenses beyond the Born approximation. We then compute in Section~\ref{sec:S3} how those post-Born terms impact the convergence skewness and take this correction into account into the large-deviation formalism in order to compute the one-point post-Born convergence PDF in Section~\ref{sec:PDF}. This corrected formalism is successfully tested against numerical simulations. Finally, Section~\ref{sec:conclusion} wraps up.

\section{Post-Born corrections to the weak lensing fields}
\label{sec:postborn}

In order to eventually derive the post-Born corrections to the skewness of the convergence field (see Section~\ref{sec:S3}), let us first recall here the Sachs equation, the formalism to add perturbations on top of a flat spacetime metric, how to get the effective projected curvature along the null geodesics responsible for the lensing effect, before expressing the weak lensing fields (that is to say the deformation matrix) at second order in perturbation theory. By doing so, we will get an explicit formula for the so-called lens-lens coupling and geodesic deviation that appear at this order. In all derivations throughout the paper, we use Einstein's summation convention and natural units where $c = H_0 = 1$.

\subsection{Sachs equation}

The weak lensing formalism aims at computing the so-called 2x2 deformation matrix $\mathcal{D}_{a b}$ which links the infinitesimal separation between two nearby null geodesics $\xi^{a}$ and the vectorial angle seen by the observer between the two geodesics $\theta_{O}^{a}$. If one notes $a(z)$ the cosmological scale factor, it is expressed as
\begin{equation}
    \xi^{a}= a(z) \mathcal{D}_{a b} \theta_{O}^{b}.
    \label{fun2}
\end{equation}
Note that the vectors $    \xi^{a}$ and $ \theta_{O}^{a}$ are two-dimensional since expressed in the basis of the \textit{screen} that the observer looks at. The evolution equation for $\mathcal{D}_{a b}$ is the optical Sachs equation (re-derived in Appendix~\ref{app::sachs} where some references are also given) which reads \begin{equation}
    \frac{\mathrm{d}^{2}}{\mathrm{d} \lambda^{2}} a(z) \mathcal{D}_{a b}=a(z)\mathcal{R}_{a}^{\ c}\mathcal{D}_{c b},
\label{Sachs22}
\end{equation}
where $\mathcal{D}_{a b}(0) = 0$ and ${\rm d}\mathcal{D}_{a b}/{\rm d}\lambda(0) = \delta_{ab}$ with $\delta_{ab}$ the identity matrix. These conditions express that the geodesics are focused at the observer and that spacetime near the observer is Euclidean. $\lambda$ parametrises the geodesics in such a way that we assume the same value at the observer, $\lambda_O = 0$, and the optical tidal tensor $\mathcal{R}_{a b}$ is expressed from the Riemann curvature contracted with null vectors $k^\nu$ and screen basis-vectors $n_{a}^{\mu}$ and $n_{b}^{\sigma}$,
\begin{equation}
    \mathcal{R}_{a b} \equiv R_{\mu \nu \rho \sigma} k^{\nu} k^{\rho} n_{a}^{\mu} n_{b}^{\sigma}.
    \label{rab3}
\end{equation}

\subsection{FLRW optical tidal tensor}
Up to this point no usual weak lensing approximations were made (e.g weak  fields which will be used in this section and Born approximation which only enters in the next section). We will now compute the matrix $\mathcal{R}_{ab}$ given a linearly perturbed flat FLRW metric which, given the Newtonian gauge and spherical coordinates, can be expressed as 
\begin{equation}
    \mathrm{d} s^{2}=-(1+2 \psi) \mathrm{d} t^{2}+a^{2}(t)(1-2\phi)\ [\mathrm{d} x^{2}+x^{2} \mathrm{d}\Omega^2 ],
\end{equation}
with $\mathrm{d}\Omega^2= \mathrm{d}\theta^2+\sin^2 \theta \mathrm{d}\varphi^2$.
According to the definition given in equation~(\ref{rab3}), we need to contract the Riemann curvature tensor at first order in $\phi$ and $\psi$ with the 4-momentum of the photon $k^{\mu}$ and $n_a^{\mu}$ computed at the background so that ${\mathcal R}_{ab}$ is also first order - linear - in $\phi$ and $\psi$. 
The Riemann tensor is straightforwardly computed from the metric so that only remains the need to express $k^{\mu}$ and $n_a^{\mu}$. 
This can be done noticing that the geodesics of interest are given by $\rm{d}\theta = \rm{d}\varphi = \rm{d}s = 0$ 
which gives the relation between ${\rm d}t$ and ${\rm d}x$ and thus between $k^{0} \equiv \mathrm{d} t / \mathrm{d} \lambda$ and $k^{1} \equiv \mathrm{d} x / \mathrm{d} \lambda$. $k^{\mu}$ is therefore proportional to $(-1,1/a(t),0,0)$ 
and the proportionality factor comes from the resolution of the geodesic equation. Finally we get
\begin{equation}
    k^{\mu}=\frac{1}{a(t)}(-1, 1/a(t), 0,0).
    \label{kmu2}
\end{equation}
We are relatively free to choose the vectors $n_a^{\mu}$ since only orthogonality conditions must be satisfied. We choose them proportional to $(0,0,1,0)$ and $(0,0,0,1)$ where the prefactors are again obtained with the geodesic equation. Thus, the screen basis vectors eventually read
\begin{equation}
\left\{ 
    \begin{aligned}
    n_1^{\mu} &= \frac{1}{a(t) x}(0,0,1,0)\\
    n_2^{\mu} &= \frac{1}{a(t) x \sin (\theta)}(0,0,0,1)
    \label{namu2}
\end{aligned}
.\right.
\end{equation}
Then plugging equations~(\ref{kmu2})-(\ref{namu2}) into equation~(\ref{rab3}) we obtain that at first order in $\phi$
\begin{equation}
    \mathcal{R}_{a b}=\frac{1}{a^2(t)}\left[-(\phi_{,ab}+\psi_{,ab})- 4\pi G \bar{\rho} \delta_{a b}\right],
\end{equation}
where the derivatives are taken along the transverse spatial components and where we used the fact that for a flat universe and a pressure-less fluid
\begin{equation}
    \frac{\Dot{a}^2- a\Ddot{a}}{a^2} = -\Dot{H} =  \frac{3}{2} H^2 = 4\pi G\bar{\rho}.
\end{equation}
Moreover, in a universe without anisotropic stress, the case considered here, we also have $\phi = \psi$. Given the relation between the scale factor and redshift together with the comoving Poisson equation
\begin{equation}
    \Delta \phi = 4\pi G \bar{\rho}\delta_{\rm mass},
    \label{poisson}
\end{equation}
where $\delta_{\rm mass} = (\rho - \bar{\rho})/\bar{\rho}$ is the density contrast, $\Delta$ is the three-dimensional Laplacian operator and derivatives are taken with respect to proper distances, one eventually gets -- up to a total derivative in $\lambda$ with no observational consequences --
\begin{equation}
    \mathcal{R}_{a b} = -\frac{3}{2}\Omega_m(1+z)^5\left[\delta_{ab} + \phi_{,ab}\right],
    \label{rab2}
\end{equation}
where we redefined $\phi$ such that $\frac{1}{2} \Delta \phi = \delta_{\rm mass}$. Since this result was computed by linearly perturbing the metric, the potential defined is a reflection of the density contrast computed in the linear regime. Note moreover that for our purposes, Newtonian perturbation theory will be sufficient for the description of the higher order density field and thus $\mathcal{R}_{ab}$ can be expanded following
\begin{align}
    \mathcal{R}_{ab} &= \mathcal{R}_{ab}^{(0)} + \mathcal{R}_{ab}^{(1)} + \mathcal{R}_{ab}^{(2)}+ ... \\ &=-\frac{3}{2}\Omega_m(1+z)^5\left[\delta_{ab} + \phi_{,ab}^{(1)}+ \phi_{,ab}^{(2)}+ ...\right]
\end{align}
where $\frac{1}{2}\Delta \phi^{(n)} = \delta_{\rm mass}^{(n)}$. Moreover we will suppose that $\mathcal{D}_{ab}$ can also be expanded with respect to $\mathcal{R}_{ab}$. Thus equation~(\ref{Sachs22}) gives a hierarchy  of equations for $\mathcal{D}_{ab}^{(n)}$ that can be solved order by order.

\subsection{Deformation  matrix at second order}

The zeroth order version of equation~(\ref{Sachs22}) corresponds to the case of an homogeneous universe  for which one gets
\begin{equation}
    \frac{{\rm d}^2[a(\lambda)\mathcal{D}_0(\lambda)]}{{\rm d}\lambda^2} = -\frac{3}{2}\Omega_m(1+z)^4\mathcal{D}_0(\lambda),
\end{equation}
whose - only - solution is a well-known result of standard flat FLRW cosmology,
\begin{equation}
    \mathcal{D}_0(\lambda(z)) = \int_0^{z} \frac{dz'}{H(z')}.
    \label{angularcomoving}
\end{equation}
Going one step further, the first order equation can be written as
\begin{multline}
        \frac{{\rm d}^2[a(\lambda)\mathcal{D}_{ab}^{(1)}(\bm{\theta},\lambda)]}{{\rm d}\lambda^2} + \frac{3}{2}\Omega_m(1+z)^4\mathcal{D}_{ab}^{(1)}(\bm{\theta},\lambda) = \\ -\frac{3}{2}\Omega_m(1+z)^4\mathcal{D}_0(\lambda) \phi^{(1)}_{,ab}(\bm{\theta},\lambda),
        \label{order1}
\end{multline}
with $\left.\mathcal{D}_{ab}^{(1)}\right|_{z=0} \! \! \! \! \! \!= 0$ and $\left.{{\rm d}\mathcal{D}_{ab}^{(1)}}/{{\rm d} \lambda}\right|_{z= 0} \! \! \! \! \! \! = 0$. Note that the associated homogeneous equation is the zeroth order one and thus the Green function $H(z) \mathcal{D}_0(z',z)$ is essentially the angular comoving distance between the redshift of the source and some other redshift. We thus obtain the first order expression of the deformation matrix by integrating this Green function on the source term
\begin{multline}
    \mathcal{D}_{ab}^{(1)}(\bm{\theta},z) = -\frac{3}{2}\Omega_m \int_0^z \frac{{\rm d}z'}{H(z')} \mathcal{D}_0(z',z) \mathcal{D}_0(z') \\ (1+z') \phi^{(1)}_{,ab}(\bm{\theta},z').
    \label{D1}
\end{multline}
Getting to this result requires expressing equation~(\ref{order1}) in terms of redshift derivatives coming from d$\lambda = (1+z)^{-2} H(z)^{-1}{\rm d}z$ which parameterises the unperturbed geodesic. This makes use of the Born approximation  that will be corrected for in the paragraphs below.

The second order equation can be similarly written as
\begin{multline}
    \frac{\mathrm{d}^{2}\left[a(z) \mathcal{D}_{i j}^{(2)}(\boldsymbol{\theta}, z)\right]}{\mathrm{d} \lambda^{2}}-a(z) \mathcal{R}_{i k}^{(0)}(z) \mathcal{D}_{k j}^{(2)}(\boldsymbol{\theta}, z)=  -\frac{3}{2} \Omega_{m}\\(1+z)^{4} \phi_{i j}^{(2)}(\boldsymbol{\theta}, z) \mathcal{D}_{0}(z) +a(z) \mathcal{R}_{i k}^{(1)}(\boldsymbol{\theta}, z) \mathcal{D}_{k j}^{(1)}(\boldsymbol{\xi}, z),
\end{multline}
where we once again recognise the zeroth order equation in the left-hand side so that the deformation matrix at second order eventually reads
\begin{multline}
    \mathcal{D}_{ab}^{(2)}(\bm{\theta},z) = -\frac{3}{2}\Omega_m \int_0^z \frac{{\rm d}z'}{H(z')} \mathcal{D}_0(z',z) \mathcal{D}_0(z') (1+z') \\ \left[\phi^{(2)}_{,ab}(\bm{\theta},z') + \frac{\phi^{(1)}_{,ac}(\bm{\theta},z')\mathcal{D}_{cb}^{(1)}(\bm{\theta},z')}{\mathcal{D}_0(z')}\right].
    \label{D2}
\end{multline}
Note that the structure of the first term in equation~(\ref{D2}) is very similar to the first order result obtained in equation~(\ref{D1}), that is a weighted superposition of second derivatives of the gravitational potential taken at the same order. This contribution exactly corresponds to summing the result of independent single thin lenses along the line-of-sight using geometrical optics. 
This is thus exact only at first order. The structure of the second term in equation~(\ref{D2}) -- a double integral -- accounts for couplings between the lenses: the effect of one particular lens along the line-of-sight depends on the effect of all the lenses upstream. In other words, the impact parameter of a photon passing through one lens depends on the source but also on the other lenses the photon passed by. More quantitatively, one can also understand this second term as a generic optical geometric correction whose form appears when one rigorously treats the case of more than one thin lens. This contribution is one of the two post-Born corrections we intend to account for, the other one comes from the fact that lenses themselves are lensed and therefore are not exactly at the angular position they would have been in the absence of lenses. This could have been directly taken into account if we for example knew how to link the perturbed geodesic parameter to redshift and then had followed the exact same steps as presented here. This will be investigated in the paragraph named "geodesic deviation" of Section~\ref{sec:kappa} below. Before that, let us first go from the deformation matrix to the convergence field.

\subsection{Convergence at second order}
\label{sec:kappa}
We now define as usual the \textit{convergence} field as the isotropic deformation of the bundle of light 
\begin{equation}
    \kappa = 1 - \frac{1}{2}{\rm Tr}\left[\frac{\mathcal{D}_{ab}}{\mathcal{D}_0}\right].
\end{equation}
 From equations~(\ref{D1}) and (\ref{D2}), noting that $\kappa$ can also be expanded with respect to the density field and has no (0) component, we get the convergence at first and second order 
\begin{multline}
    \kappa^{(1)}({\bm \theta}, z) = \frac{3}{2} \Omega_m  \!\!\int_0^{z} \!\!\frac{{\rm d}z'}{H(z')} \frac{\mathcal{D}_0(z',z) \mathcal{D}_0(z')}{\mathcal{D}_0(z)} \\\times (1+z') \delta_{{\rm mass}}^{(1)}({\bm \theta}, z'),
    \label{kappa1}
\end{multline}

\begin{multline}
    \kappa^{(2)}({\bm \theta}, z) = \frac{3}{2} \Omega_m \!\!\int_0^{z}\!\! \frac{{\rm d}z'}{H(z')} \frac{\mathcal{D}_0(z',z) \mathcal{D}_0(z')}{\mathcal{D}_0(z)} (1+z') \delta_{{\rm mass}}^{(2)}({\bm \theta}, z') \\+ \kappa^{(2)}_{{\rm corr.1}}({\bm \theta}, z)
    \label{kappa2}
\end{multline}
where $\kappa^{(2)}_{{\rm corr.1}}({\bm \theta}, z)$ gathers contributions from the coupling of lenses that are second order in the field (there is none at first order as stated above for the deformation matrix) and will be explicitly given below. Before, discussing each term one by one, let us first note that
the weighting factor along the line-of-sight defined by
\begin{equation}
    \omega(z',z) = \frac{3}{2}\Omega_m \frac{\mathcal{D}_0(z',z) \mathcal{D}_0(z')}{\mathcal{D}_0(z)} (1+z'),
\end{equation}
usually called lensing efficiency function or lensing kernel, characterises the efficiency of the lenses along the line-of-sight.
Note also that we use $\Delta \phi^{(n)} = 2\delta_{\rm mass}^{(n)}$ to get equations~(\ref{kappa1}) and (\ref{kappa2}) relying on the fact that the term $\phi_{,33}$, where the direction 3 is along the line-of-sight, cancels out in the integration. This corresponds to a plane-parallel approximation, that is to say only waves perpendicular to the line-of-sight contribute to lensing, which holds in the small-angle cases and is usually referred to as the Limber approximation. 
Note that at the scales of interest here (namely below the degree), this approximation is extremely accurate (see for example \cite{Lemos17} for an illustrative plot for the power spectrum). 
Let us now discuss all terms that contribute to the convergence field at second order, one by one.

\paragraph*{Born approximation with independent lenses}
First, as shown in the previous derivation, the convergence $\kappa$ gets a contribution at every order from the integration along the line-of-sight of the same order density contrast. This contribution tends to be the dominant term. Thus it is common to approximate the convergence by the density field projected along the line-of-sight such that
\begin{equation}
    \kappa({\bm \theta}, z) = \int_0^{z} \frac{{\rm d}z'}{H(z')} \omega(z',z) \delta_{{\rm mass}}({\bm \theta}, z'), 
\end{equation}
where $\delta_{{\rm mass}}({\bm \theta}, z')$ is the non-linear matter density contrast. 
However we have seen that this approximation is strictly valid only in the linear regime and some corrections arise at second order and beyond. Thus only in this approximation is the convergence \textit{potential}, and can therefore be seen as the Laplacian of some projected gravitational potential. As a consequence, only in this approximation schemes in the spirit of \cite{KaiserSquires} to reconstruct convergence maps from observed shear maps can be applied. Though the convergence could still also be reconstructed, for example, from magnification of sources, formally taking into account corrections to this approximation alters our ability to actually get to an "observed" convergence. Going beyond the Born and independent lenses approximation, let us now discuss the post-Born corrections arising at this order.

\paragraph*{Lens-lens coupling}

At second order the correction accounts for the couplings of lenses along the line-of-sight and can be written as\footnote{Let us re-emphasize here that there is an implicit summation over the repeated indices and that  indices $a$ and $b$ run only on the transverse directions due to the use of the Limber approximation.}
\begin{multline}
    \kappa^{(2)}_{{\rm corr.1}}({\bm \theta}, z) = - \frac{1}{2} \int_0^z \frac{{\rm d}z'}{H(z')} \omega(z',z) \int_0^{z'} \frac{{\rm d}z''}{H(z'')} \omega(z'',z')\\ \times \phi^{(1)}_{,ab}({\bm \theta},z')\phi^{(1)}_{,ab}({\bm \theta},z'')
    \label{corr1}
\end{multline}
which comes directly from equations~(\ref{D2}) and (\ref{kappa2}). This term is exactly the same as the second term of equation~(\ref{D2}) already discussed, but now expressed for the convergence rather than the full deformation matrix. Being typically negative, this corrective term typically contributes to gaussianising the field and lessens the importance of quantities beyond the two-point correlation function. 

\begin{figure}
    \centering
    \includegraphics[width = \columnwidth]{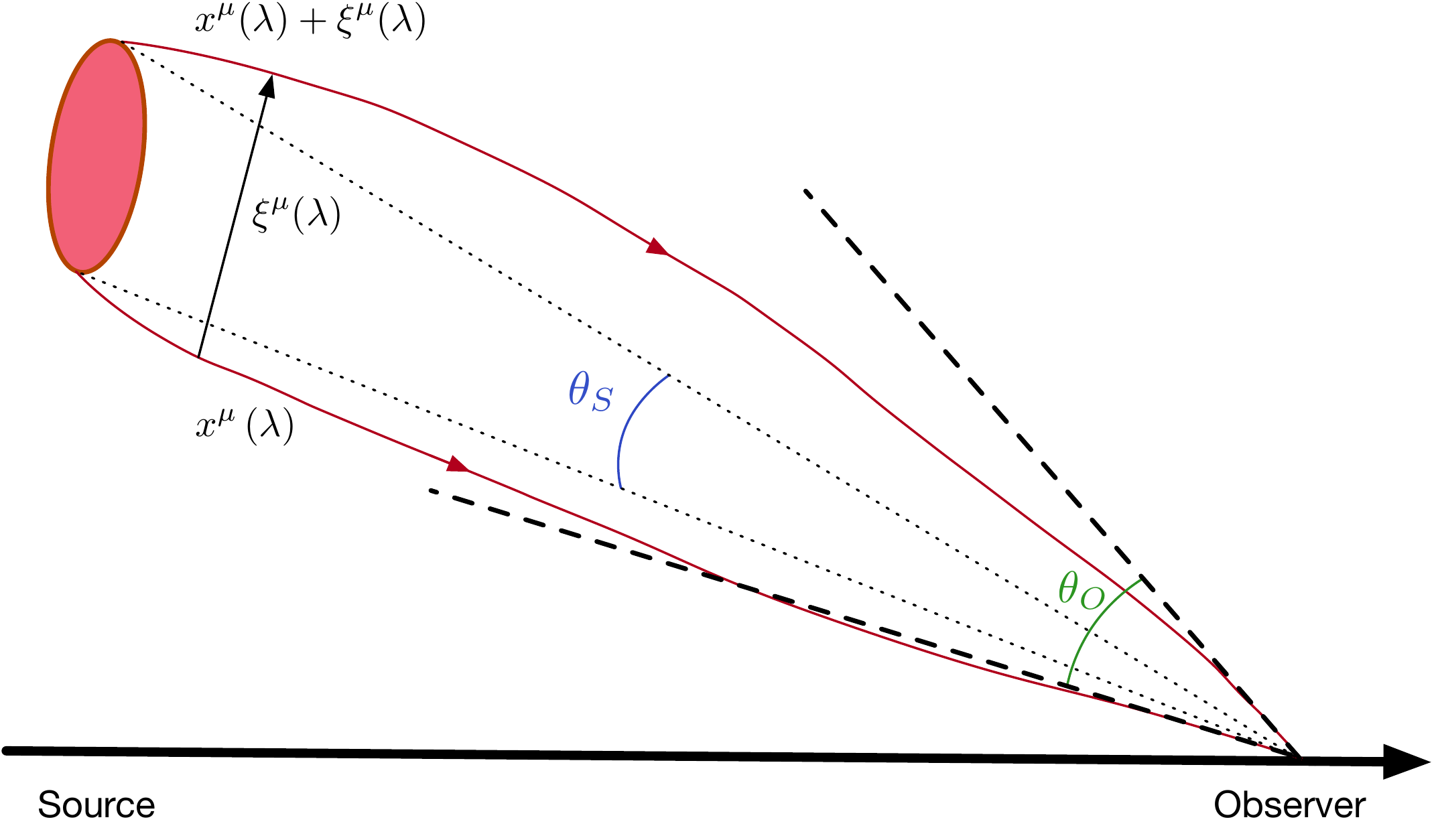}
    \caption{Schematic view of possible trajectories of photons coming from a source to an observer. Quantities of interest are labelled as in the main text.}
    \label{schema}
\end{figure}

\paragraph*{Geodesic deviation}
\label{gd}
In the derivation of the convergence,
we made use of the Born approximation, and therefore integrate along the unperturbed line-of-sight. However, this approximation is not valid in a lumpy universe and one needs to account for additional couplings between lenses due to the perturbations of the trajectories of photons. Hence, let us now include in the line-of-sight integrations the fact that the potential and density are not to be taken at angular position ${\bm \theta}$, as in equation~(\ref{kappa1}), but at ${\bm \theta} + {\rm d} {\bm \theta}$ where ${\rm d}{\bm \theta}$ is the deflection induced by the foreground lenses. To compute its value, first note that equation~(\ref{fun2}) shows that $\mathcal{D}_{ij}/\mathcal{D}_0$ is the Jacobian of the local transformation from the real (source) angle to the apparent (image) angle. 
Then considering a lens equation of the type $\theta^{\rm S}_i = \theta^{\rm O}_i + {\rm d}\theta_i\left(\theta^{\rm O}\right)$, where we follow notations of Fig.~\ref{schema}, provides us with a direct link between the deflection angle ${\rm d}\theta_i\left(\theta^{\rm O}\right)$ we are looking for and $\mathcal{D}_{ij}$. Using equation~(\ref{D1}), one gets at first order
\begin{equation}
    {\rm d}\theta_a^{(1)}({\bm \theta},z) = - \int_0^z \frac{{\rm d}z'}{H(z')} \frac{\omega(z',z)}{\mathcal{D}_0(z')} \phi^{(1)}_{,a}(\bm{\theta},z').
\end{equation}
Now taylor expanding the density contrast
\begin{equation}
    \delta_{\mathrm{mass}}^{(1)}(\boldsymbol{\theta}+{\rm d} \boldsymbol{\theta}, z) \approx \delta_{\mathrm{mass}}^{(1)}(\boldsymbol{\theta}, z)+\nabla_{\boldsymbol{\theta}} \delta^{(1)}_{\mathrm{mass}}(\boldsymbol{\theta}, z) \cdot {\rm d} \boldsymbol{\theta}, 
    \label{dth}
\end{equation}
and combining equations~(\ref{dth}) and (\ref{kappa1}) yields an expression for the leading order correction to the Born approximation which is second order for the convergence (unsurprising since it accounts for couplings between lenses and therefore cannot appear at first order)
\begin{multline}
    \kappa^{(2)}_{{\rm corr.2}}({\bm \theta}, z) = - \frac{1}{2} \int_0^z \frac{{\rm d}z'}{H(z')} \omega(z',z) \int_0^{z'} \frac{{\rm d}z''}{H(z'')} \omega(z'',z') \\ \frac{\mathcal{D}_0(z')}{\mathcal{D}_0(z'')} \phi^{(1)}_{,aab}({\bm \theta},z')\phi^{(1)}_{,b}({\bm \theta},z''),
    \label{corr2}
\end{multline}
where derivatives are still taken with respect to transverse spatial components. Note that similarly to (\ref{corr1}), this correction also tends to gaussianise the field. This is natural as the accumulation of successive almost independent kicks along the line-of-sight can only tend to blur the field and makes it seem more Gaussian. Finally note that the derivation of lensing quantities we presented here is done in many references using the so-called Fermat's principle. This approach leads to the same equations and is thus equivalent to our formalism\footnote{The complete derivation proposed in this paper therefore justifies the use of Fermat's principle in this context.}. This is demonstrated in appendix~\ref{Sec:Fermat} which also contains the generalisation of the deformation matrix at second order given by equation~(\ref{D2}) with post-Born corrections (see equation~(\ref{eq:PBdef})).
Note that this equation could also straightforwardly give the flexion after differentiation with respect to the apparent angular position.

\section{Post-Born corrections to the skewness}
\label{sec:S3}

\subsection{Definition of the skewness}

Given the second-order corrections to the convergence obtained in the previous section, let us now compute the corrections induced on the skewness of the convergence defined as
\begin{equation}
    S_{3, \kappa} = \frac{\left\langle \kappa^3 \right\rangle_c}{\left\langle \kappa^2 \right\rangle_c^2} = \frac{\left\langle \kappa^3 \right\rangle}{\left\langle \kappa^2 \right\rangle^2},
\end{equation}
since moments and cumulants (connected part of the moments, denoted with the subscript $_c$) are equal at second and third order for a random variable with zero mean $\langle \kappa \rangle = 0$. The leading order term to the third moment is given by
\begin{equation}
    \begin{aligned}
    \left\langle\kappa^{3}\right\rangle &=\left\langle\left(\kappa^{(1)}+\kappa^{(2)}+\ldots\right)^{3}\right\rangle \\ &=\left\langle\left(\kappa^{(1)}\right)^{3}\right\rangle+ 3\left\langle\left(\kappa^{(1)}\right)^{2} \kappa^{(2)}\right\rangle+\ldots  \\ 
    &=  3\left\langle\left(\kappa^{(1)}\right)^{2} \kappa^{(2)}\right\rangle+\ldots
    \end{aligned}
\end{equation}
because the linear convergence does not admit a skewness term if no primordial non-Gaussianities are assumed (as we do in this work). As such, the leading order correction to the skewness induced by the second order post-Born corrective terms is given by 
\begin{equation}
\label{eq:S3}
    S_{3, \kappa}^{\rm corr} = 3 \frac{\left\langle\left(\kappa^{(1)}\right)^{2} \kappa^{(2)}_{\rm corr}\right\rangle}{\left\langle\left(\kappa^{(1)}\right)^{2}\right\rangle^2}.
\end{equation}
Finally, we are interested in the filtered convergence field. Following \cite{Barthelemy19} we use an angular top-hat window of radius $\theta$ which 
reads in Fourier space $W\left[k_{\perp} \mathcal{D}_{0}(z) \theta\right] $
where 
$k_{\perp}$ is the norm of {\bf k} in the transverse directions and 
\begin{equation}
    W(l) = 2\frac{J_1(l)}{l},
\end{equation}
with $J_1$ the first order Bessel function of the first kind.
Note that the linear variance of the convergence field  filtered with an opening angle of $\theta$
which enters equation~(\ref{eq:S3})
can be computed from equation~(\ref{kappa1}) as was done in \cite{Bernardeau1997}. In particular, the projection formula for cumulants given in equation (14) of  \cite{Barthelemy19} yields
\begin{multline}
    \left\langle\left(\kappa^{(1)}_{\theta}\right)^{2}\right\rangle = \frac{1}{2 \pi} \int_0^{z} \frac{{\rm d}z'}{H(z')} \omega(z',z)^2 D_+(z')^2 \\ \int_0^{+ \infty} {\rm d}k_{\perp} k_{\perp} P_l(k_{\perp}) W(k_{\perp}\mathcal{D}_0(z')\theta)^2 ,
    \label{variance}
\end{multline}
where $D_+(z)$ is the linear growth factor and $P_l(k)$ the linear matter power spectrum.

\subsection{Quantification of the post-Born corrections}

In order to explicitly compute the post-Born corrections to the skewness, let us first express
the gravitational potential $\phi$ in Fourier space as
\begin{equation}
    \phi({\bm \theta},z) = \int \frac{{\rm d}{\bm k}}{(2 \pi)^{3/2}} \tilde{\phi}({\bm k},z) e^{{\rm i}({\bm k}_\perp \cdot {\bm \theta} + k_r \chi(z))},
\end{equation}
which can be related to the density field via a Fourier transform of equation~(\ref{poisson}) so that at first order
\begin{equation}
    \tilde{\phi}^{(1)}({\bm k},z) = \frac{2}{k^2} D_+(z) \tilde{\delta}^{(1)}_{\rm mass}({\bm k}).
\end{equation}
From there, one can easily get the terms $\phi^{(1)}_{,ab}({\bm \theta},z)\phi^{(1)}_{,ab}({\bm \theta},z')$ and $\phi^{(1)}_{,aab}({\bm \theta},z)\phi^{(1)}_{,b}({\bm \theta},z')$ entering the post-Born corrections in equations~(\ref{corr1}) and (\ref{corr2}). First, the contraction of second derivatives can be written
\begin{multline}
\label{eq:phi22}
    \phi^{(1)}_{,ab}({\bm \theta},z)\phi^{(1)}_{,ab}({\bm \theta},z') = \frac{4}{(2 \pi)^3} \!\!\int\! {\rm d} {\bm k}{\rm d} {\bm k}' \left[\frac{({\bm k}\cdot {\bm k}')^2}{k^2 k'^2} \right] e^{{\rm i}({\bm k}_\perp +{\bm k}'_{\perp})\cdot {\bm \theta}}\\ \times e^{{\rm i}( k_r \chi(z)+ k'_r \chi(z'))} D_+(z)D_+(z') \tilde{\delta}^{(1)}_{\rm mass}({\bm k})\tilde{\delta}^{(1)}_{\rm mass}({\bm k}'),
\end{multline}
where the small-angle (Limber)  approximation ${\bm k}_\perp^2 \sim {\bm k}^2$ was used. Similarly, the contraction of third and first derivatives in the geodesic deviation reads
\begin{multline}
\label{eq:phi31}
    \phi^{(1)}_{,aab}({\bm \theta},z)\phi^{(1)}_{,b}({\bm \theta},z') = \frac{4}{(2 \pi)^3} \int {\rm d} {\bm k}{\rm d} {\bm k}' \left[\frac{{\bm k}\cdot {\bm k}'}{k'^2} \right]e^{{\rm i}({\bm k}_\perp +{\bm k}'_{\perp})\cdot {\bm \theta}} \\ e^{{\rm i}( k_r \chi(z)+ k'_r \chi(z'))} D_+(z)D_+(z') \tilde{\delta}^{(1)}_{\rm mass}({\bm k})\tilde{\delta}^{(1)}_{\rm mass}({\bm k}').
\end{multline}
Equations~(\ref{eq:phi22}) and (\ref{eq:phi31}) have the same structure. We thus try and gather them using the kernel
\begin{equation}
    G({\bm k},{\bm k}',z',z'') = \frac{\mathcal{D}_0(z')}{\mathcal{D}_0(z'')}\frac{{\bm k}\cdot {\bm k}'}{k'^2} + \frac{({\bm k}\cdot {\bm k}')^2}{k^2k'^2},
\end{equation}
such that the corrective terms defined in equations~(\ref{corr1}) and (\ref{corr2}) can be jointly written as
\begin{multline}
\label{eq:kappa2PB}
    \kappa^{(2)}_{{\rm corr}}({\bm \theta}, z) = - \frac{1}{2}\frac{4}{(2 \pi)^3} \int_0^z \!\! \frac{{\rm d}z'}{H(z')} \omega(z',z) \int_0^{z'} \!\!\! \frac{{\rm d}z''}{H(z'')} \omega(z'',z')  \\ \int {\rm d} {\bm k}{\rm d} {\bm k}' G({\bm k},{\bm k}',z',z'') e^{{\rm i}(({\bm k}_\perp +{\bm k}'_{\perp})\cdot {\bm \theta} + k_r \chi(z)+ k'_r \chi(z'))} \\ D_+(z')D_+(z'') \tilde{\delta}^{(1)}_{\rm mass}({\bm k})\tilde{\delta}^{(1)}_{\rm mass}({\bm k}').
\end{multline}
From there, the computation of the post-Born correction to the third order moment is very similar to that of the usual skewness. We first make use of Wick's theorem to simplify the integrand appearing when plugging equation~(\ref{eq:kappa2PB}) into (\ref{eq:S3})
\begin{multline}
    \int {\rm d}{\bm k_1}{\rm d}{\bm k_2}{\rm d}{\bm k_3}{\rm d}{\bm k_4} \left\langle\tilde{\delta}^{(1)}_{\rm mass}({\bm k_1})\tilde{\delta}^{(1)}_{\rm mass}({\bm k_2})\tilde{\delta}^{(1)}_{\rm mass}({\bm k_3})\tilde{\delta}^{(1)}_{\rm mass}({\bm k_4}) \right\rangle \\ G({\bm k_3}, {\bm k_4},z'',z''') e^{\mathrm{i}{\bm k_3}\cdot {\bm r}''(z'')} e^{\mathrm{i}{\bm k_4}\cdot {\bm r}''(z''')} e^{\mathrm{i}{\bm k_1}\cdot {\bm r}(z)} e^{\mathrm{i}{\bm k_2}\cdot {\bm r'}(z')} = \\ \int {\rm d}{\bm k_1}{\rm d}{\bm k_2} P_l(k_1)P_l(k_2)G({\bm k_1}, {\bm k_2},z'',z''') e^{\mathrm{i}{\bm k_1}\cdot {\bm r}''(z'')} e^{\mathrm{i}{\bm k_2}\cdot {\bm r}''(z''')} \\\qquad\quad\times \left[e^{-\mathrm{i}{\bm k_1}\cdot {\bm r}(z)} e^{-\mathrm{i}{\bm k_2}\cdot {\bm r}'(z')}+ e^{-\mathrm{i}{\bm k_1}\cdot {\bm r}'(z')} e^{-\mathrm{i}{\bm k_2}\cdot {\bm r}(z)}\right] \\ + \int {\rm d}{\bm k_1}{\rm d}{\bm k_2} P_l(k_1)P_l(k_2) \left(1 - \frac{\mathcal{D}_0(z'')}{\mathcal{D}_0(z''')} \right) \\ \times e^{{\rm i}\left({\bm k_1}\cdot {\bm r}''(z'')-{\bm k_1}\cdot {\bm r}''(z''') \right)} e^{{\rm i}\left({\bm k_2}\cdot {\bm r}(z)-{\bm k_2}\cdot {\bm r}'(z') \right)},
\end{multline}
where ${\bm r}(z) = (\chi(z),{\bm \theta})$. Then filtering the field with an opening angle of $\theta$ and integrating over the radial wave vectors we get 
\begin{multline}
 \int {\rm d}{\bm k_1}{\rm d}{\bm k_2}{\rm d}{\bm k_3}{\rm d}{\bm k_4} \left\langle\tilde{\delta}^{(1)}_{\rm mass}({\bm k_1})\tilde{\delta}^{(1)}_{\rm mass}({\bm k_2})\tilde{\delta}^{(1)}_{\rm mass}({\bm k_3})\tilde{\delta}^{(1)}_{\rm mass}({\bm k_4}) \right\rangle \\ G({\bm k_3}, {\bm k_4},z'',z''') e^{\mathrm{i}{\bm k_3}\cdot {\bm r}''(z'')} e^{\mathrm{i}{\bm k_4}\cdot {\bm r}''(z''')} e^{\mathrm{i}{\bm k_1}\cdot {\bm r}(z)} e^{\mathrm{i}{\bm k_2}\cdot {\bm r'}(z')} = \\
    (2 \pi)^2 \int {\rm d}^2{\bm k_1}{\rm d}^2{\bm k_2} P_l(k_1)P_l(k_2)G({\bm k_1}, {\bm k_2},z,z')
   \\ W(|{\bm k_1}\mathcal{D}_0(z'')+{\bm k_2}\mathcal{D}_0(z''')|\theta) W(k_1\mathcal{D}_0(z)\theta) W(k_2\mathcal{D}_0(z')\theta) \\ \delta_D(\chi(z'')-\chi(z))\delta_D(\chi(z''')-\chi(z')) + \text{perm}(z \leftrightarrow z').
\end{multline}
Making the change of variable ${\bm k_1}\mathcal{D}_0(z'') \rightarrow {\bm \ell_1}$ and ${\bm k_2}\mathcal{D}_0(z''') \rightarrow {\bm \ell_2}$ allows us to express the correction to the third order moment as
\begin{multline}
    \left\langle\left(\kappa^{(1)}\right)^{2} \kappa^{(2)}_{\rm corr}\right\rangle_\theta = - \frac{4}{(2 \pi)^4} \int_0^{z} \!\!\! \frac{{\rm d}z'}{H(z')} \omega(z',z)^2 \int_0^{z'} \!\! \frac{{\rm d}z''}{H(z'')} \\ \omega(z'',z) \omega(z'',z') D_+(z')^2D_+(z'')^2 \! \int \!\!\! \frac{{\rm d}{\bm \ell_1} {\rm d}{\bm \ell_2}}{\left[\mathcal{D}_0(z')\mathcal{D}_0(z'')\right]^2} P_l\left(\frac{\ell_1}{\mathcal{D}_0(z')}\right) \\ P_l\left(\frac{\ell_2}{\mathcal{D}_0(z'')}\right) H({\bm \ell_1},{\bm \ell_2}) W(\ell_1\theta)W(\ell_2\theta)W(|{\bm \ell_1}+{\bm \ell_2}|\theta),
    \label{k3}
\end{multline}
where \footnote{As expected, the kernel $H$ does not have any residual redshift dependence (which is a consequence of ) and is null for $\ell_1+\ell_2=0$.}
\begin{equation}
    H({\bm \ell_1},{\bm \ell_2}) = \frac{{\bm \ell_1}\cdot {\bm \ell_2}}{\ell_2^2} + \frac{({\bm \ell_1}\cdot {\bm \ell_2})^2}{\ell_1^2\ell_2^2}.
    \label{H}
\end{equation}

Using the following properties \citep{Bernardeau1995} of the 2D top-hat window function where $\varphi$ is the angle between the two wave-modes ${\bm \ell}_1$ and ${\bm \ell}_2$
\begin{equation}
    \int_{0}^{2 \pi} \mathrm{d} \varphi W\left(\left|\boldsymbol{\ell}_{1}+\boldsymbol{\ell}_{2}\right|\right)\left[1-\cos ^{2}(\varphi)\right]=\pi W\left(\ell_{1}\right) W\left(\ell_{2}\right),
    \label{prop1}
\end{equation}
\begin{multline}
    \int_{0}^{2 \pi} \mathrm{d} \varphi W\left(\left|\boldsymbol{\ell}_{1}+\boldsymbol{\ell}_{2}\right|\right) \left[1+\cos (\varphi) \frac{\ell_{1}}{\ell_{2}}\right] \\ =2 \pi W\left(\ell_{2}\right)\left[W\left(\ell_{1}\right)+\frac{\ell_{1}}{2} W^{\prime}\left(\ell_{1}\right)\right],
    \label{prop2}
\end{multline}
and plugging them in equation~(\ref{k3}) finally yields
\begin{multline}
    \left\langle\kappa^3_{\rm corr}\right\rangle_\theta = - \frac{48}{(2 \pi)^2} \int_0^{z_s} \frac{{\rm d}z'}{H(z')} \omega(z',z_s)^2 \int_0^{z'} \frac{{\rm d}z''}{H(z'')} \omega(z'',z_s) \\ \omega(z'',z') D_+(z')^2D_+(z'')^2 \int \frac{{\rm d}{\ell_1} {\rm d}{\ell_2}}{\left[\mathcal{D}_0(z')\mathcal{D}_0(z'')\right]^2}  P_l\left(\frac{\ell_1}{\mathcal{D}_0(z')}\right) \\ P_l\left(\frac{\ell_2}{\mathcal{D}_0(z'')}\right) \frac{J_1(\ell_1 \theta ) (J_0(\ell_1 \theta )-J_2(\ell_1 \theta )) J_1(\ell_2 \theta ){}^2}{\ell_2 \theta ^3}, 
    \label{finalk3corr}
\end{multline}
where the subscript $_s$ now denotes the source plane. This notation is introduced since no distribution of sources will be taken into account. The generalisation is nevertheless straightforward. The integral in equation~(\ref{finalk3corr}) can then be computed numerically.

\begin{figure}
    \centering
    \includegraphics[width=\columnwidth]{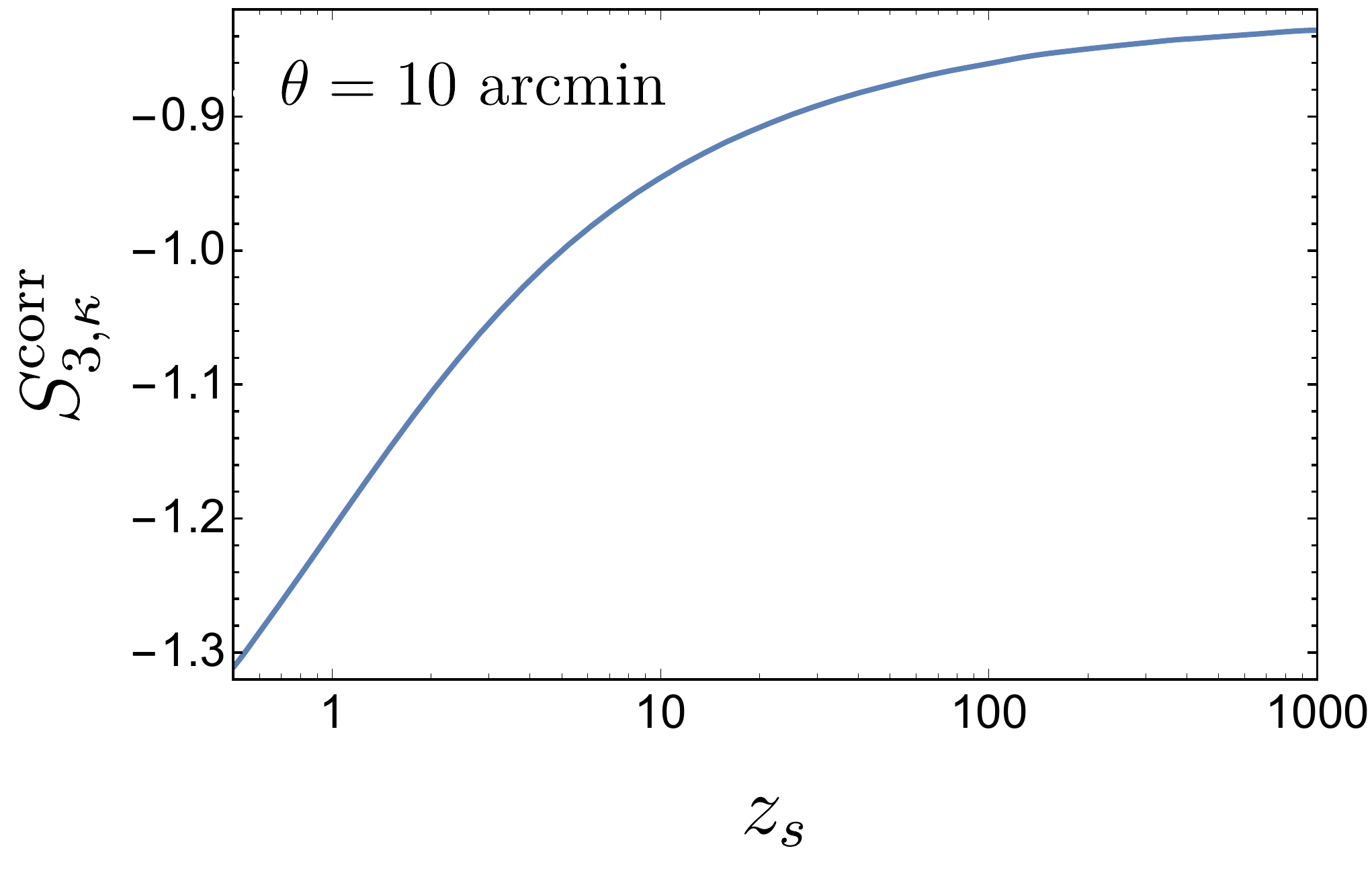}
    \caption{Leading order correction to the skewness induced by lens-lens couplings and geodesic deviation. The field is filtered in top-hat window of angular radius $\theta = 10$ arcmin and we plot the evolution of the correction with respect to the source redshift.}
    \label{PlotSkewness}
\end{figure}

\begin{figure}
    \centering
    \includegraphics[width=\columnwidth]{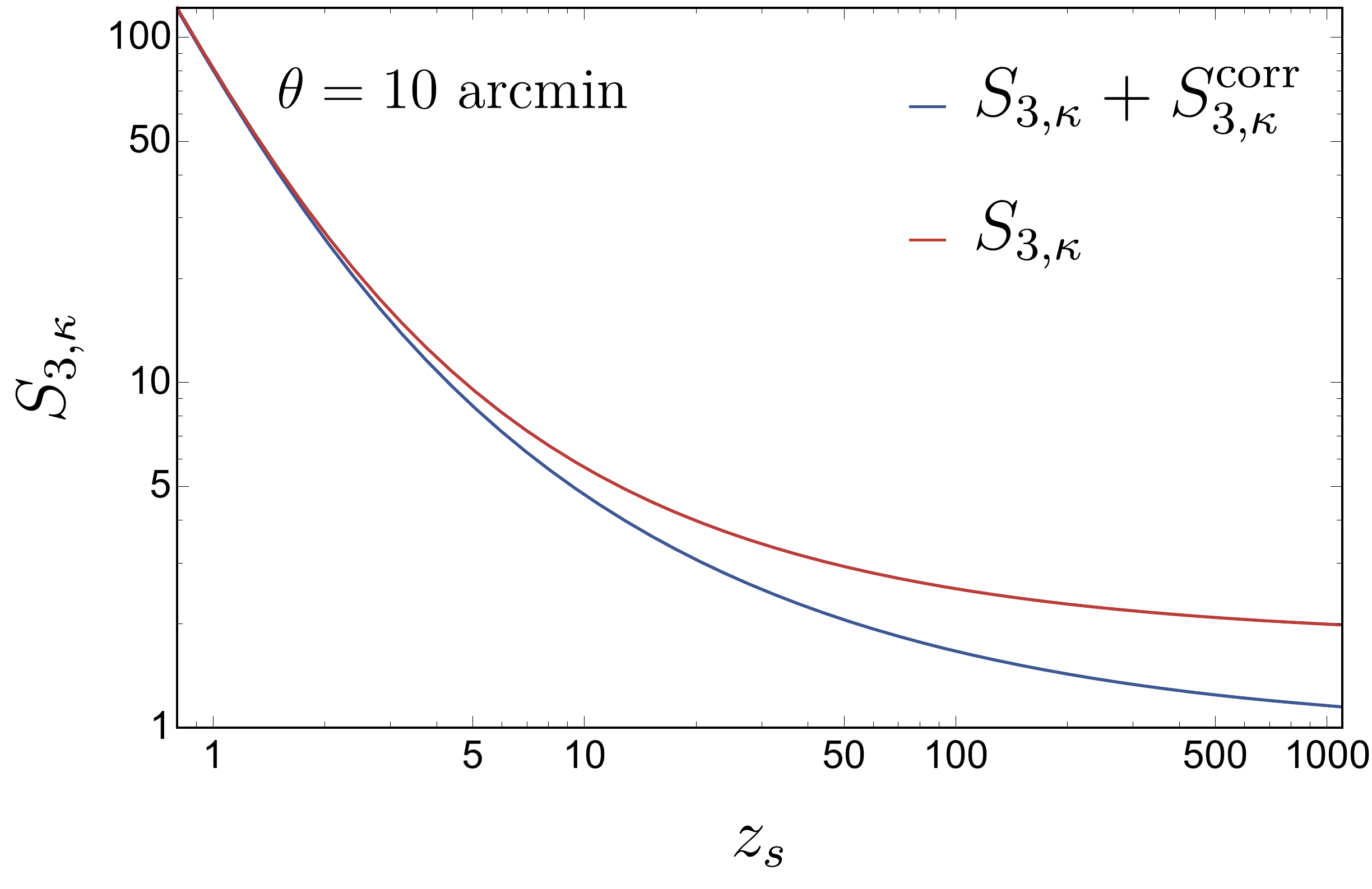}
    \caption{Leading order skewness with (blue) and without (red) considering the correction induced by lens-lens couplings and geodesic deviation. The field is filtered with a top-hat window of angular radius $\theta = 10$ arcmin and we plot the evolution of the skewness with respect to the source redshift.}
    \label{comparison}
\end{figure}

For an opening angle $\theta =$ 10 arcmin we obtain the correction on the skewness due to lens-lens coupling and geodesic deviation shown in Fig.~\ref{PlotSkewness} (as a function of the source redshift). Note that this is the correction to the so-called tree-order skewness since we only consider the first non zero term beyond the linear regime. The correction, for this scale and up to high source redshifts, is shown to be very small given the tree-order order skewness displayed in Fig.~\ref{comparison} (blue solid line), it barely reaches a few percents of the signal at low redshifts. As expected, the correction becomes relatively more important as the source redshift increases. Indeed, the correction appears to be relatively constant in time whereas the tree-order skewness value roughly decreases as $z_s^{-1.35}$. The effect is thus proven to be more important at high redshift, especially in the context of CMB lensing for which the post-Born corrections have the same amplitude as the signal itself and tend to make the field even  more Gaussian (with a skewness going roughly from 2 to 1 for the aperture of 10 arcmin used here).

\begin{figure}
    \centering
    \includegraphics[width = \columnwidth]{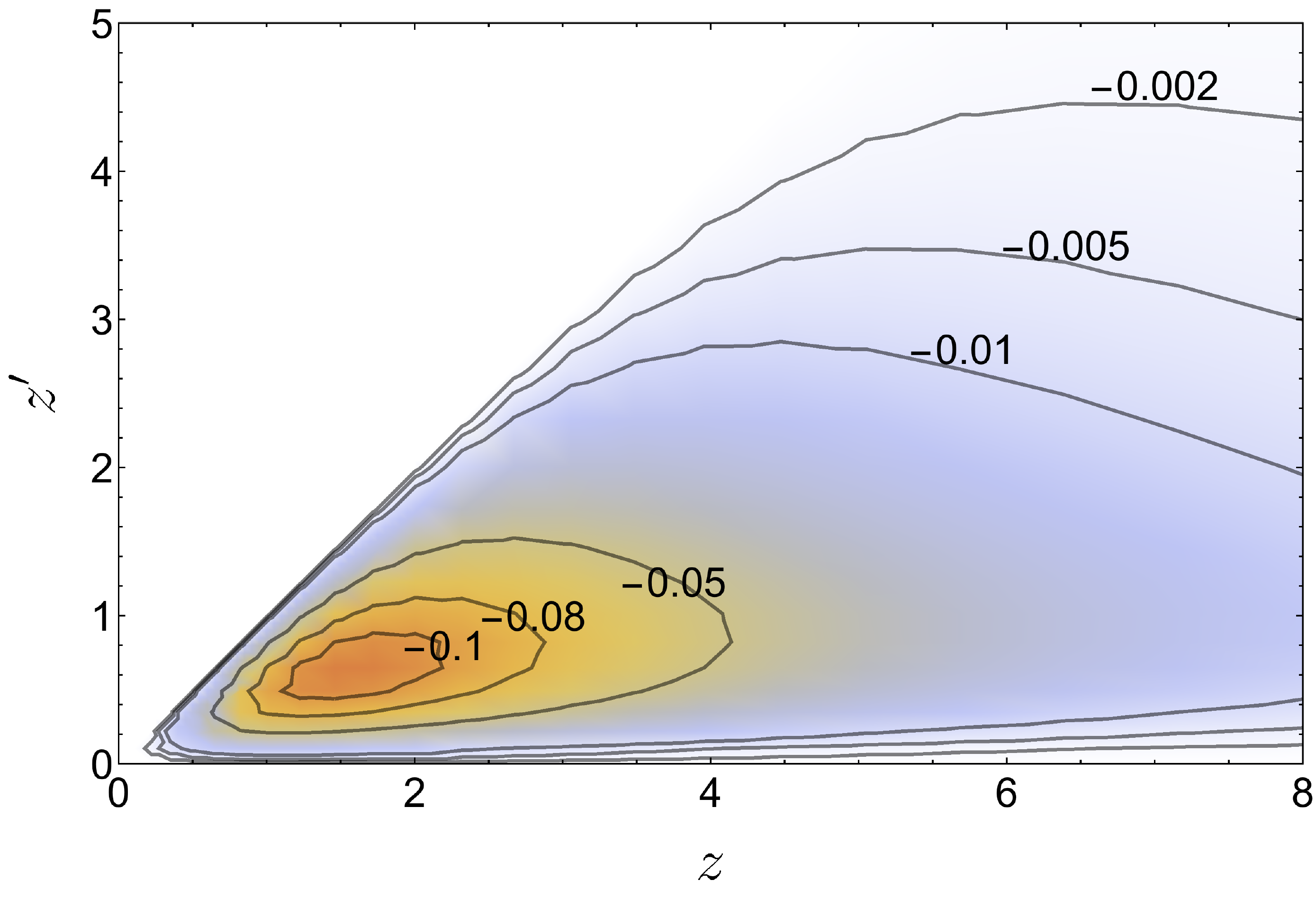}
    \caption{Contribution to the correction of the skewness from couplings between lenses (including the so-called lens-lens coupling and geodesic deviation) at redshift $z$ and $z'$ for the CMB convergence field and an aperture $\theta=10$ arcmin.}
    \label{density}
\end{figure}

We also show in Fig.~\ref{density}, in the case of a source at $z_s = 1100$ and an opening angle of $\theta = 10$ arcmin, the contribution to the skewness due to the couplings between the lenses at redshift $z$ and $z'$. This contribution is null when the lenses are at the same redshift or one of them is at the observer or the source i.e $z'=0$ or $z=z_s$ and peaks at relatively low redshift $(z,z')\sim (1.5,0.7)$.

\section{PDF with post-Born corrections}
\label{sec:PDF}
Let us now incorporate the calculation of the skewness with post-Born corrections performed in the previous section into the one-point PDF of the convergence field.
This PDF has been modelled in the past using Edgeworth expansion and improved more recently with large-deviation theory \citep{BernardeauValageas,Munshi00,Valageas1,Valageas2,Munshi14,paolo,Barthelemy19} but no post-Born corrections were included so far. 
The strategy to adopt in this respect is not trivial. After a quick summary of the large-deviation formalism to predict the one-point PDF of cosmic fields, we first propose a naive approach which may work when the corrections are small but fails when they become more important because of the so-called cumulant problem. 
Finally, we investigate a new method in line with the very spirit of large-deviation theory, which proves to successfully predict the convergence PDF especially for large post-Born corrections as is the case in the context of CMB lensing.

\subsection{One-point PDF without \nth{2} order corrections}
\label{formalism}

Let us first remind the very basics of large deviation theory in the context of cosmic structure formation. A random variable $\rho$ -- {\red admitting a PDF ${\mathcal P}_{\rho}(\rho,\sigma^2)$  where $\sigma^2$ is called the inverse driving parameter which is the variance of the field in our case --} is said to follow a large deviation principle if the following limit exists
\begin{equation}
    \Psi_{\rho}(\rho) = - \lim_{\sigma^2 \rightarrow 0} \sigma^2 \log({\mathcal P}_{\rho}(\rho,\sigma^2)).
    \label{LDP}
\end{equation}
It therefore defines $\Psi_{\rho}(\rho)$, the rate function characterising the exponential decay of the distribution of values for $\rho$ in the limit where the {\red inverse driving parameter (variance in our case)} of the field is small.

The existence of a large deviation principle for $\rho$ implies
that the scaled cumulant generating function (SCGF) $ \varphi_{\rho}$ is given through Varadhan's theorem as the Legendre-Fenchel transform of the rate function
\begin{equation}
    \varphi_{\rho}(y) = \sup_{\rho} \,[y\rho - \Psi_{\rho}(\rho)],
    \label{varadhan}
\end{equation}
where the Legendre-Fenchel transform reduces to a simple Legendre transform when $\Psi_{\rho}$ is convex. In that case, 
\begin{equation}
    \varphi_{\rho}(y) =  y\rho - \Psi_{\rho}(\rho),
    \label{Legendre}
\end{equation}
where $\rho$ is a function of $y$ through the following stationary condition
\begin{equation}
    y = \frac{\partial \Psi_{\rho}}{\partial \rho}.
    \label{stationnary}
\end{equation}
{\red Let us also remind that this SCGF is a re-scaling, when the inverse driving parameter goes to zero, of the usual cumulant generating function (CGF) $\phi_\rho$ defined as
\begin{equation}
    \phi_\rho(y)= \log \left[\int_{-\infty}^{+ \infty} e^{y \rho} {\mathcal P}_\rho(\rho,\sigma^2) {\rm d}\rho\right],
    \label{phidef}
\end{equation}
and we have 
\begin{equation}
    \varphi_\rho(y)= \lim_{\sigma^2 \rightarrow 0} \sigma^2\phi_\rho\left(\frac{y}{\sigma^2}\right).
    \label{varphiCGF}
\end{equation}
Thus plugging equation~(\ref{LDP}) in equation~(\ref{phidef}) one can check that Varadhan's theorem is a saddle point approximation which becomes exact in the limit of zero variance. As such, one key point of the application of large deviation theory in the context of cosmic structure formation is to extrapolate these asymptotic results to finite values of the variance, thus left as a free parameter since the theory does not strictly speaking enable its computation. Hence, one wants to get an estimate of the CGF -- dropping the limit and inversing equation~(\ref{varphiCGF}) -- from the exact computation of the SCGF and finally go back to the field PDF.}

Another consequence of the large-deviation principle, very useful in the cosmological context, 
is the so-called contraction principle.
This principle states that if we consider a random variable $\tau$ related to $\rho$ through the continuous map $f$ then its rate function can be computed as
\begin{equation}
    \Psi_{\rho}(\rho) = \inf_{\tau:f(\tau) = \rho} \Psi_{\tau}(\tau).
    \label{contraction}
\end{equation}
This formula is called the contraction principle because $f$ can be many-to-one. 
In other words, there might be many $\tau$ such that $\rho = f(\tau)$, 
in which case we are {\it contracting}  information about the rate function of $\tau$ down to $\rho$. 
In physical terms, this formula is interpreted by saying that an improbable fluctuation of $\rho$ is brought about by the most probable of all improbable fluctuations of $\tau$.
Hence, the rate function of the late-time density field can be computed from the initial conditions if the most likely mapping between the two is known, 
that is if one is able to identify the leading field configuration that will contribute to this infimum. 

In spherically symmetric configurations (for instance mean cosmic densities in spheres of a given radius), one could conjecture \citep{Valageas} that the most likely mapping between initial and final conditions is spherical collapse for which an accurate parametrisation is given by
\begin{equation}
    \zeta(\tau) = \left(1 - \frac{\tau}{\nu} \right)^{-\nu},
    \label{collapse}
\end{equation}
where $\nu$ may be chosen so as to reproduce the tree-order skewness $\nu=21/13$.

Note that spherical collapse is equivalent to tree-order perturbation theory \citep{Bernardeau1992} but we get all cumulants at once (there is no truncation at some order in the hierarchy of cumulants). This property is essential to capture the tails of the distribution. This is to be contrasted with Edgeworth-like expansions that truncate the expansion at a given order and are therefore very inaccurate in the tails.

Hence, starting from Gaussian initial conditions\footnote{Primordial non-Gaussianities could also straightforwardly be accounted for in this formalism as shown by \cite{NonGaussianities}.} one can compute i) the initial rate function which is nothing but a quadratic function of the linear density contrast, ii) the late-time rate function using the contraction principle given by equation~(\ref{contraction}) and iii) the SCGF from equation~(\ref{varadhan}). 
From this SCGF, one can then compute the cumulant generating function for infinitesimal slices of density and when neglecting deviations from the Born approximation and lens-lens couplings one can apply the projection formula in \cite{Barthelemy19} to eventually obtain the SCGF of the convergence field
\begin{equation}
    \varphi_{\kappa}(y) \!=\!\!\! \int_0^{\chi_s} \!\!\!\!\!{\rm d}\chi \, \frac{\left\langle{\kappa^{(1)}}^2\right\rangle_{\theta}}{\left\langle{\delta^{(1)}}^2_{\rm slice}\right\rangle}  \varphi^{\rm slice}\!\!\!\left(\!\omega(z_s,z)\frac{\left\langle{\delta^{(1)}}^2_{\rm slice}\right\rangle}{\left\langle{\kappa^{(1)}}^2\right\rangle_{\theta}}  y,z\!\right)\!,
    \label{phiproj}
\end{equation}
which then translates {\red (inversing equation~(\ref{phidef}))} into the one-point convergence PDF via an inverse Laplace transform
\begin{equation}
    {\mathcal P}_\kappa(x) = \int_{-i\infty}^{+i\infty} \frac{{\rm d}y}{2\pi i} \, \text{exp}\left(-y x + \frac{\varphi_\kappa(y \sigma_\kappa^2)}{\sigma_\kappa^2}\right).
    \label{eq::laplace}
\end{equation}
In equation~(\ref{eq::laplace}), {\red and in accordance with the fact that we extend the theory to finite values of the inverse driving parameter,} the non-linear variance of the convergence $\sigma_\kappa^2$ is obtained by some other mean, for example from a Halofit power spectrum \citep{Halofit}. For more details, we refer the readers to \cite{Barthelemy19}.

Using this approach boils down to constructing a PDF based on all the tree-order cumulants of the convergence. One could for example see it as an infinite Edgeworth expansion in which all cumulants are computed at tree order in perturbation theory and as such, follow a well-known hierarchy as they scale as $S_n\propto\sigma^{2(n-1)}$\citep{Fry84}. Therefore, when using this method, it makes sense to compute the tree-order correction to the skewness arising from deviations to the Born approximation, at least in regimes where these corrections are significant (notably at high redshift, see Fig.~\ref{comparison}), but less so for higher order cumulants since next-to-leading contributions to the skewness are of the same order as the tree-order kurtosis.

Let us emphasise that the construction from large deviations makes sense from a mathematical point of view since it allows us to construct a PDF in a rigorous sense. Indeed, as the CGF is computed from a simple mapping of an initial Gaussian random variable before being inverse Laplace transformed, the output is indeed a function that has all the good properties of a PDF namely being positive and normalised.
This large-deviation PDF is also meaningful from a physical point of view since everything it encodes has a physical origin (tree-order perturbation theory and not a phenomenological prescription). 
Strictly speaking, this theory is exact in the limit of zero variance. The extrapolation to small non zero values is then performed without prior knowledge on the range of validity one can reach. Fortunately, it has been shown to probe accurately the mildly non-linear regime up to rather surprisingly small scales (few megaparsecs corresponding to a typical amplitude of density fluctuations of $\sigma \lesssim 1$),
relying on numerical simulations \citep{saddle,Barthelemy19}.
As a consequence, corrections arising from post-Born effects need only be computed for the tree-order skewness and included as such in the formalism.

\subsection{A "large deviationesque" approach to the post-Born convergence  PDF}
\label{deviationesque}

\begin{figure}
    \centering
    \includegraphics[width = \columnwidth]{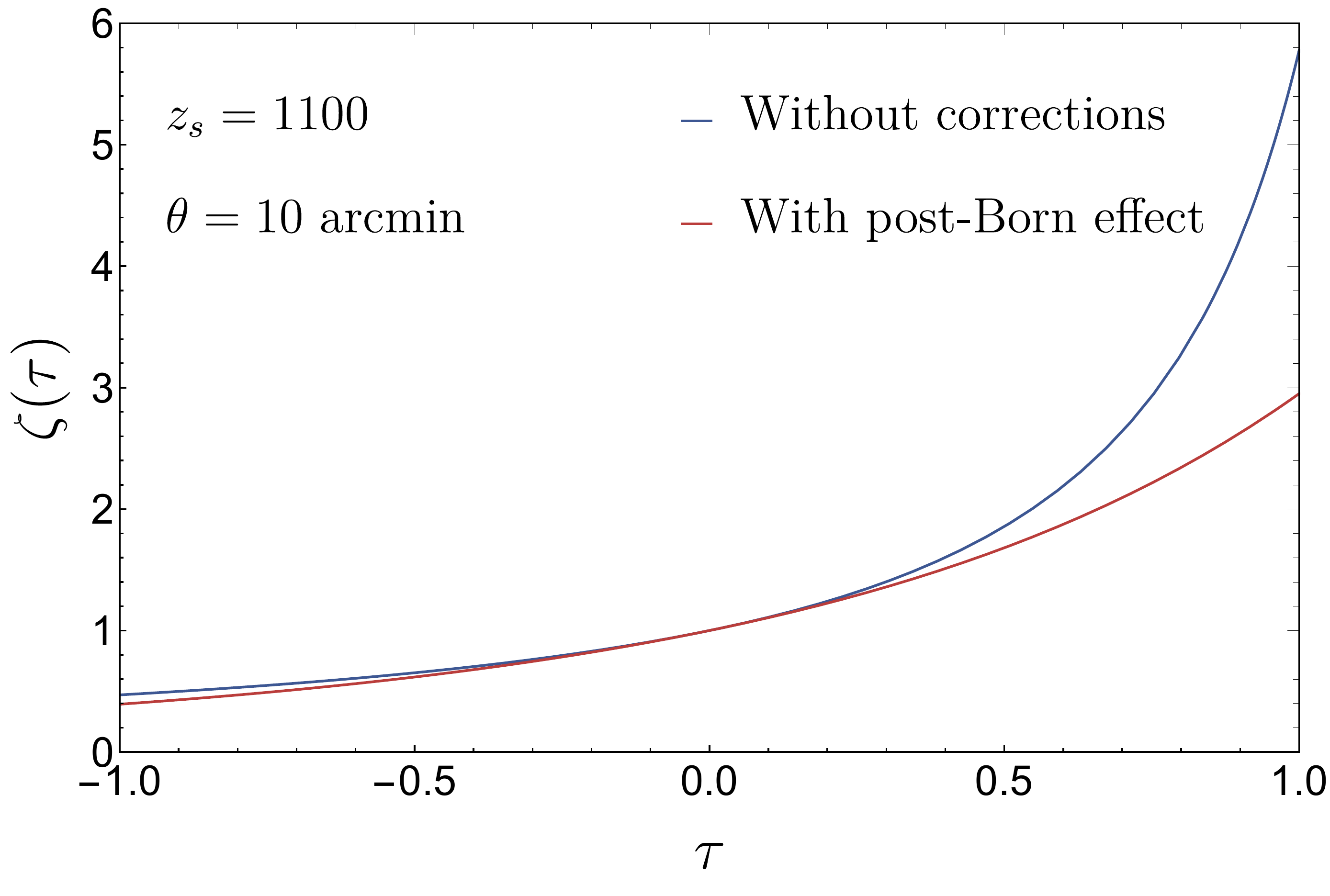}
    \caption{Modified mapping of a slice to account for post-Born corrections to the CMB convergence skewness (red) compared to the uncorrected spherical collapse dynamics (blue) given in equation~(\ref{collapse}). The corrections are introduced by changing the mapping globally and for a source redshift $z_s = 1100$ and an opening angle $\theta = 10$ arcmin ($\nu$ goes from 1.4 to 6.7).}
    \label{changemapping}
\end{figure}

{\red As was shown in the previous section, the physics of this large deviation approach of cosmic structure formation is entirely contained in the mapping from the field initial conditions to its late-time counterpart and thus any modification that one would want to incorporate should only be at the level of the mapping itself. Appendix~\ref{fast_dirty} shows how a more naive approach where the cumulants are directly modified in the late-time SCGF to incorporate corrections on the skewness is not a well-defined mathematical construction and induces unphysical behaviour. We therefore stick to the strategy mentioned before, namely a modification of the mapping itself. Note that this modification of the mapping should be done with care so as not to introduce any spurious critical points where the new rate function ceases to be convex -- in the sense that they would for example emerge from a truncated mapping series expansion --, in which case the shape of the PDF could be greatly affected.
}
Since the effect of the post-Born corrections on the skewness is subdominant for the convergence with respect to the total collapse dynamics in the cone, we might {\red thus} want to consider a modification of the implicit "conical" collapse driven by the parameter $\nu$. In this case, we keep the functional form of the 2D spherical collapse for each slice and find an overall effective $\nu$ modified for each of them. For a source redshift $z_s = 1100$ and an opening angle $\theta = 10$ arcmin for which the post-Born corrections are important, we plot in Fig.~\ref{changemapping} how the effective dynamics is affected. This change of effective mapping will also impact the value of the higher order cumulants (kurtosis and beyond) but in a meaningful, or at least mathematically consistent, way since there are now deduced at the level of the new mapping and so that issues (negative values, mis-normalisation) described in the appendix \ref{fast_dirty} are avoided. {\red Moreover, since this new construction also respects the cumulant hierarchy we can safely expect that the higher-order cumulants will at least have the good order of magnitude except maybe for highly non-linear regimes if this hierarchy does not apply.}

\subsection{Application to CMB lensing}

We showed in Figs.~\ref{PlotSkewness} and \ref{comparison} that the Post-Born corrections in the context of weak lensing experiments with sources of low redshift, $z_s \leq 3-4$, will not be of the utmost importance
and it was already showed in \cite{Barthelemy19} that those corrections do not, at the level of the PDF, impact the quality of the prediction especially when one makes use of a nulling strategy \citep{Nulling}. 
However, since the value of the tree-order correction to the skewness is nearly constant and close to $S_{3,\kappa}^{\rm corr} \approx -1$ and the (uncorrected) tree-order convergence skewness roughly decreases as $\approx z_s^{-1.35}$, depending on the precision one may need there is always a regime for which it might be mandatory to include the correction to the PDF. 

Here we will focus on CMB lensing ($z_s = 1100$) where the correction is of the order of the uncorrected skewness and thus plays an important role. Following the conclusion of Section~\ref{deviationesque}, let us globally modify the spherical collapse parameter so as to reproduce the corrected convergence skewness, plug this new parametrisation in the formalism described in Section~\ref{formalism} and thus compute the one-point PDF of the convergence field taking into account the post-Born corrections.
We will then compare the computed PDF at source redshift 1100 for an opening angle of $\theta = 10$ arcmin to the one extracted from lensing simulated maps (for which the measured standard deviation is $\sigma_{\kappa}\sim 0.036$). 

\subsubsection{Simulated maps}
\begin{table}
    \centering
    \bgroup
    \def\arraystretch{1.5}
    \begin{tabular}{|c|c|c|c|c|c|c|}
   \hline
    $\Omega_m$ & $\Omega_{\Lambda}$ & $\Omega_{cdm}$ & $\Omega_b$ & h & $\sigma_8$ & $n_s$  \\
    \hline
    0.279 & 0.721 & 0.233 & 0.046 & 0.7 & 0.82 & 0.97 \\
    \hline
    \end{tabular}
    \egroup
    \caption{Cosmological parameters used to run the simulations used in this paper.}
    \label{table1}
\end{table}

The full-sky simulated maps were generated by \cite{Simulation} who generated 108 full-sky gravitational lensing simulation data sets performing multiple-lens plane ray-tracing through high-resolution cosmological N-body simulations. A system of nested cubic simulation boxes have been prepared to reproduce the mass distribution in the Universe and placed around a fixed vertex representing the observer’s position. They have been evolved in a periodic cosmological N-body simulation following the gravitational evolution of dark matter particles without baryonic processes where initial conditions have been based on the second-order Lagrangian perturbation theory. The number of particles for each box was $2048^3$, making the mass and spatial resolutions better for the inner boxes. It was checked that the matter power spectra agreed with theoretical predictions of the revised Halofit and ray-tracing was performed using the public code {\sc graytrix} which follows the standard multiple-lens plane algorithm in spherical coordinates using the {\sc healpix} algorithm. 

The data sets include full-sky convergence maps from redshifts $z= 0.05$ to 5.3 at intervals of 150~Mpc~$h^{−1}$ comoving radial distance and are freely available for download\footnote{\url{http://cosmo.phys.hirosaki-u.ac.jp/takahasi/allsky_raytracing/}}. It also includes convergence maps at the redshift of the CMB which were generated making full use of the N-Body simulation up to redshift $\sim 7$ and generating a density field of the earlier epochs based on linear power spectrum calculated using the Code for Anisotropies in the Microwave Background ({\sc camb}, \cite{camb}). The ray-tracing scheme was then applied in the same fashion as for the other lensing maps. 
Note that although the density field at epochs beyond redshift 7 is very close to the linear regime for the scales we consider, the method used by \cite{Simulation} inevitably introduces some inaccuracies in the non-linearities of the density field and a fortiori the convergence field. However, we checked using tree-order perturbation theory that the error induced on the non linear convergence skewness at redshift 1100 and opening angle of 10 arcmin is only $\sim2$\%. This can for example be seen in Fig.~\ref{contributions}. To the best of our knowledge these full-sky simulations are nonetheless state-of-the-art. 
The adopted cosmological parameters are consistent with the WMAP-9 year result as shown in Table \ref{table1}. 
The pixelization of the full-sky maps follows the {\sc healpix} ring scheme with resolution of about 0.86~arcmin. They were convolved with a top-hat angular window function of the desired radius and we measure the non-linear variance directly in the maps. Remember that this non-linear variance is a free parameter of the theory and used as input in equation~(\ref{eq::laplace}).

Finally the error bars on the measured PDFs and higher order cumulants are estimated via the dispersion (standard deviation) amongst the 23 different realisations we considered. Note that the number 23 of realisations used to get accurate measurements was chosen after seeing that adding more realisations does not affect the values nor the size of the error bars (from 10 to 23 realisations, the mean is shifted typically by one percent and the error bar changes by a few percents in the displayed range of opening angles). 

\subsubsection{Comparison between predicted PDF and simulations}
\begin{figure}
    \centering
    \includegraphics[width =\columnwidth]{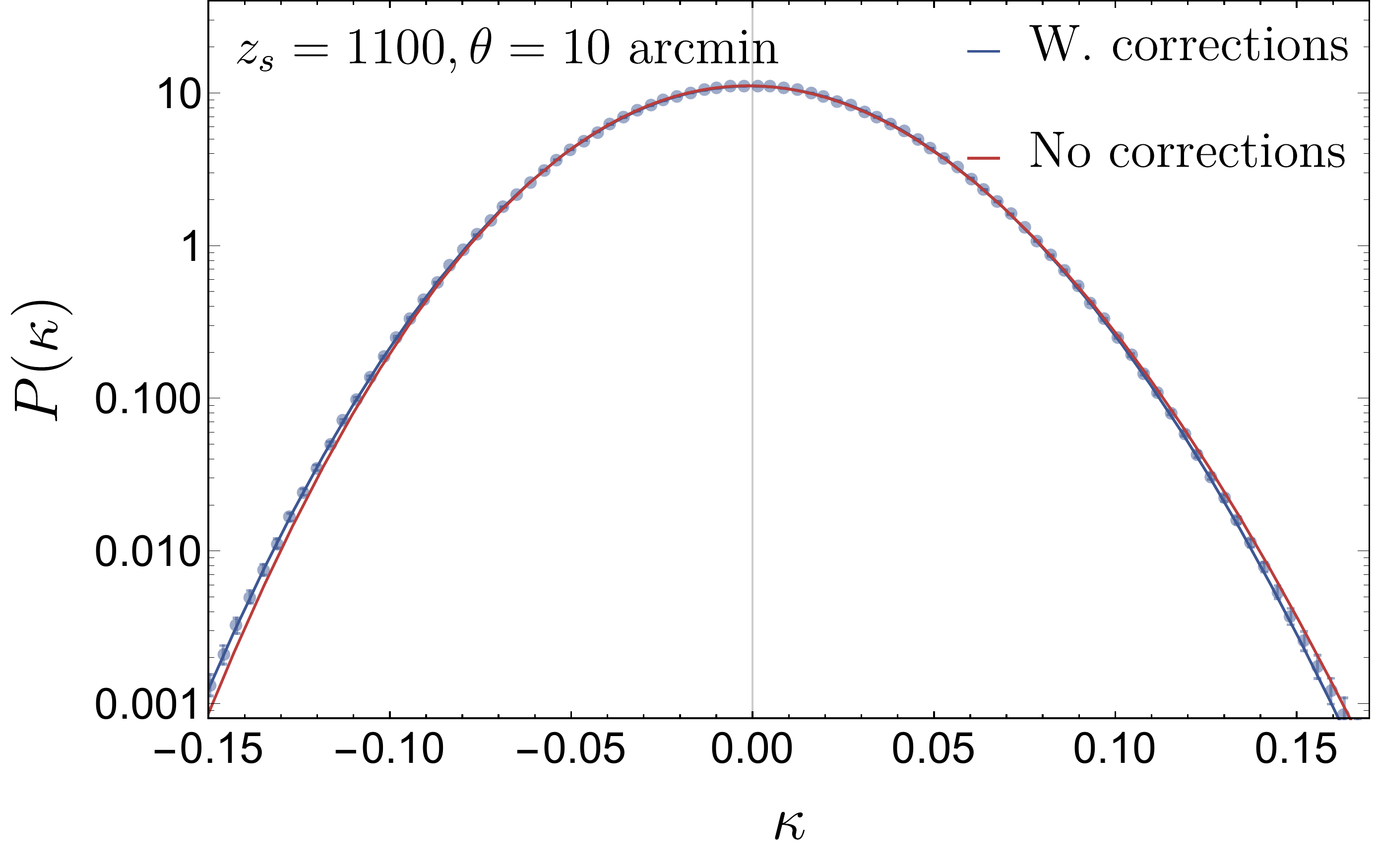}
    \includegraphics[width =\columnwidth]{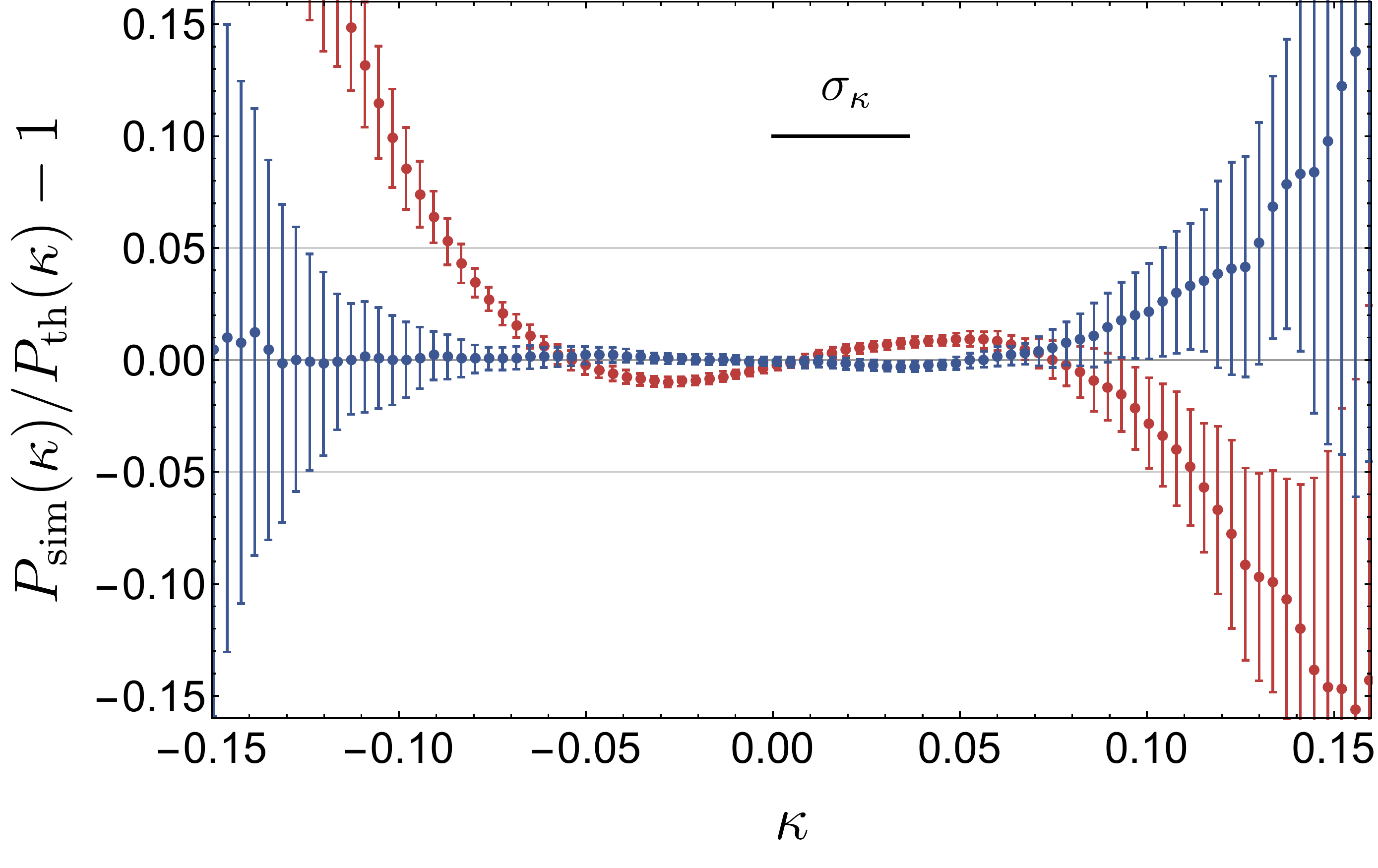}
    \caption{One point PDF of the weak lensing convergence for an opening angle of 10~arcmin as labelled. The source redshift is fixed here to $z_s$ = 1100. Solid lines display the LDT predictions given by equation~(\ref{eq::laplace}) with (blue) and without (red) implementation of post-Born corrections while the measurements on the simulated sky is shown with error bars obtained from the standard deviation between 23 realisations. Top panel: PDF in log scale to better display the tails. Bottom panel: residuals of the simulated data compared to the prediction.}
    \label{pdf}
\end{figure}

Fig.~\ref{pdf} -- the main result of this paper -- shows the predicted PDF compared to the one measured in the simulations. The first thing worth noting is that the simulated cosmic variance is relatively large as can be seen from the size of the error bars for both the tails of the PDF and the higher-order cumulants on their own\footnote{Note that large-deviation theory provides us with a way to compute the expected error bars including both the effect of shot noise and cosmic variance, we refer the reader to \cite{2016MNRAS.460.1598C} for more details.}. Nevertheless the need for going beyond Born-approximation is clear when one looks at the residuals of Fig.~\ref{pdf} (red points) that significantly shows tens of percents deviations between the simulation and the theoretical prediction without post-Born corrections, manifestly driven by an order one error on the skewness (as the S-shape suggests, being a well-known $H_3$ modulation\footnote{The $H_3$ (third order Hermite polynomial) shape of the residuals is typical of an incorrect skewness, the dominant term that needs to be corrected for in the cumulant generating function and eventually leading to the theoretical PDF. This $H_3$ polynomial notably appears as the leading correction to the Gaussian distribution in an Edgeworth expansion and is modulated by the skewness term. }).  
This observation strengthens our previous argument stating that taking into account the correction to the skewness is sufficient to get an accurate PDF. This can be quantitatively assessed by following the prescription described in Sect.~\ref{deviationesque}.
Once post-Born corrections are accounted for (blue solid line), the agreement between the theoretical formalism and the measurement in the simulation is very good within the error bars in the range of convergence probed that is to say in the $\sim 5 \sigma_{\kappa}$ region around the mean. For negative values of the convergence, there is no sign of any deviation from the prediction and a subpercent accuracy is reached. On the other hand, some hints of departures is seen in the high (positive) convergence tail for which more accurate simulations would be needed (at least for $\kappa$ above $\sim 2 \sigma_{\kappa}$). 

Overall, it is very reassuring and satisfying to see that theoretical predictions and simulations agree within a percent in a large range of values for the convergence, from $\sim -5 \sigma_{\kappa}$ to $2 \sigma_{\kappa}$, a regime where both theory and simulations have not been tested in details. For higher values of the convergence field, it is difficult to draw any conclusion at this stage with regards to either the accuracy of the prediction or the validity of the simulated maps. {\red Indeed though it is clear that the theory tends to systematically under-predict the expectation values of the PDF in the high tail, an effect which comes from the fact that the skewness itself is under-predicted with respect to its simulated expectation value, given the quite sizeable dispersion (cosmic variance displayed as the error bars) one gets between the realisations, the fact that the simulation was not extensively tested with non-Gaussian statistics or the fact that full-sky CMB lensing simulation is still something rather hard to perform we cannot exclude that some systematic comes from the simulation, nonetheless being state-of-the-art, itself. In any case this paper boils down to constructing a tree-order convergence PDF with some post-Born correction, it is thus not excluded that one might in the future have to go to higher order to get a more accurate prediction if needed.}

\begin{figure}
    \centering
    \includegraphics[width =\columnwidth]{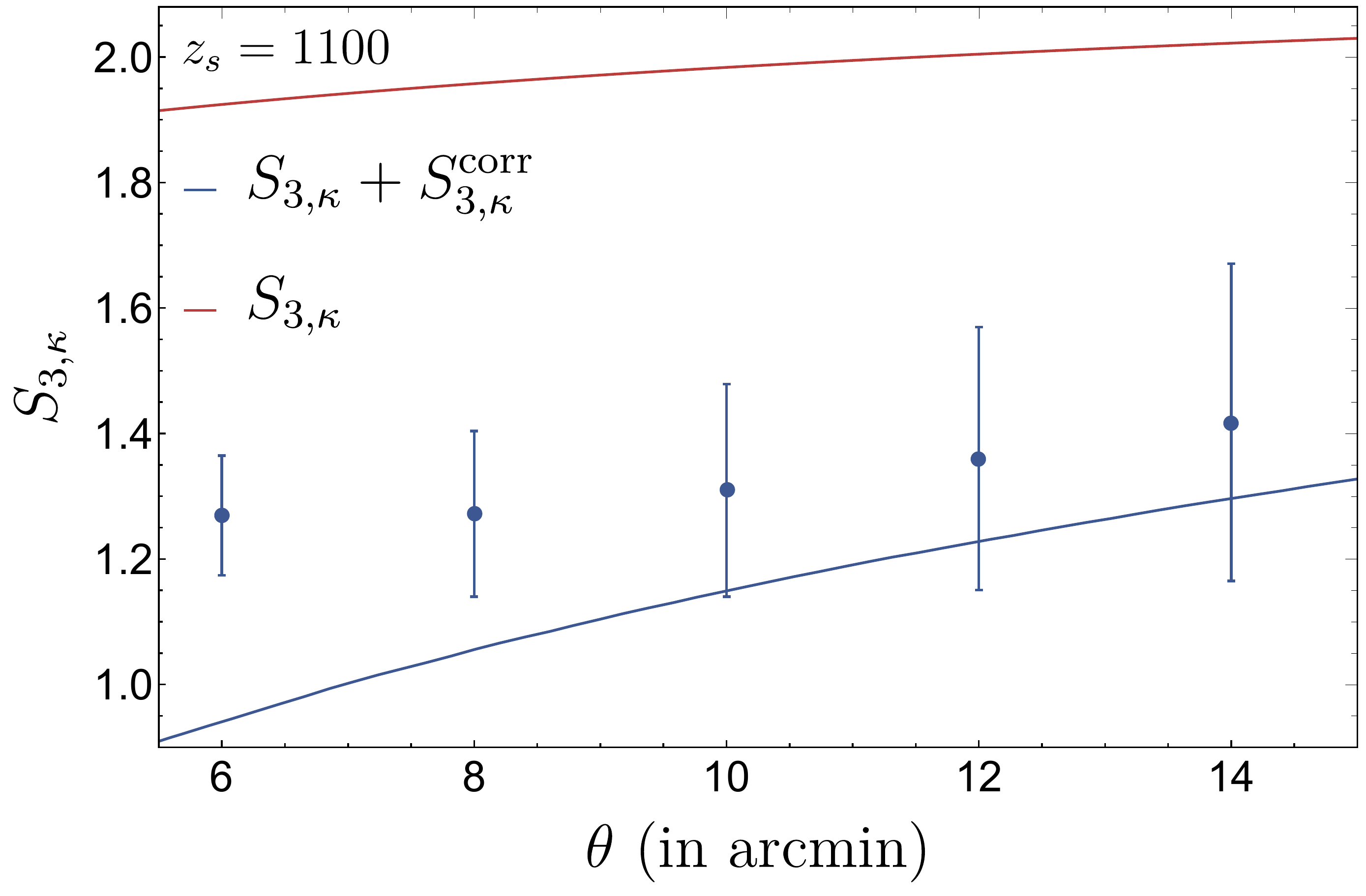}
    \caption{Convergence skewness of the CMB with respect to the opening angle of the top-hat smoothing applied. In red (resp. blue) is the skewness evolution as computed from large deviation theory without (resp. with) taking into account post-Born corrections. The measured points are from the 23 simulated maps (mean of the 23 realisations) described in the main text and error bars represent their standard deviation around the mean. 
    }
    \label{S3angle}
\end{figure}

\begin{figure}
    \centering
    \includegraphics[width =\columnwidth]{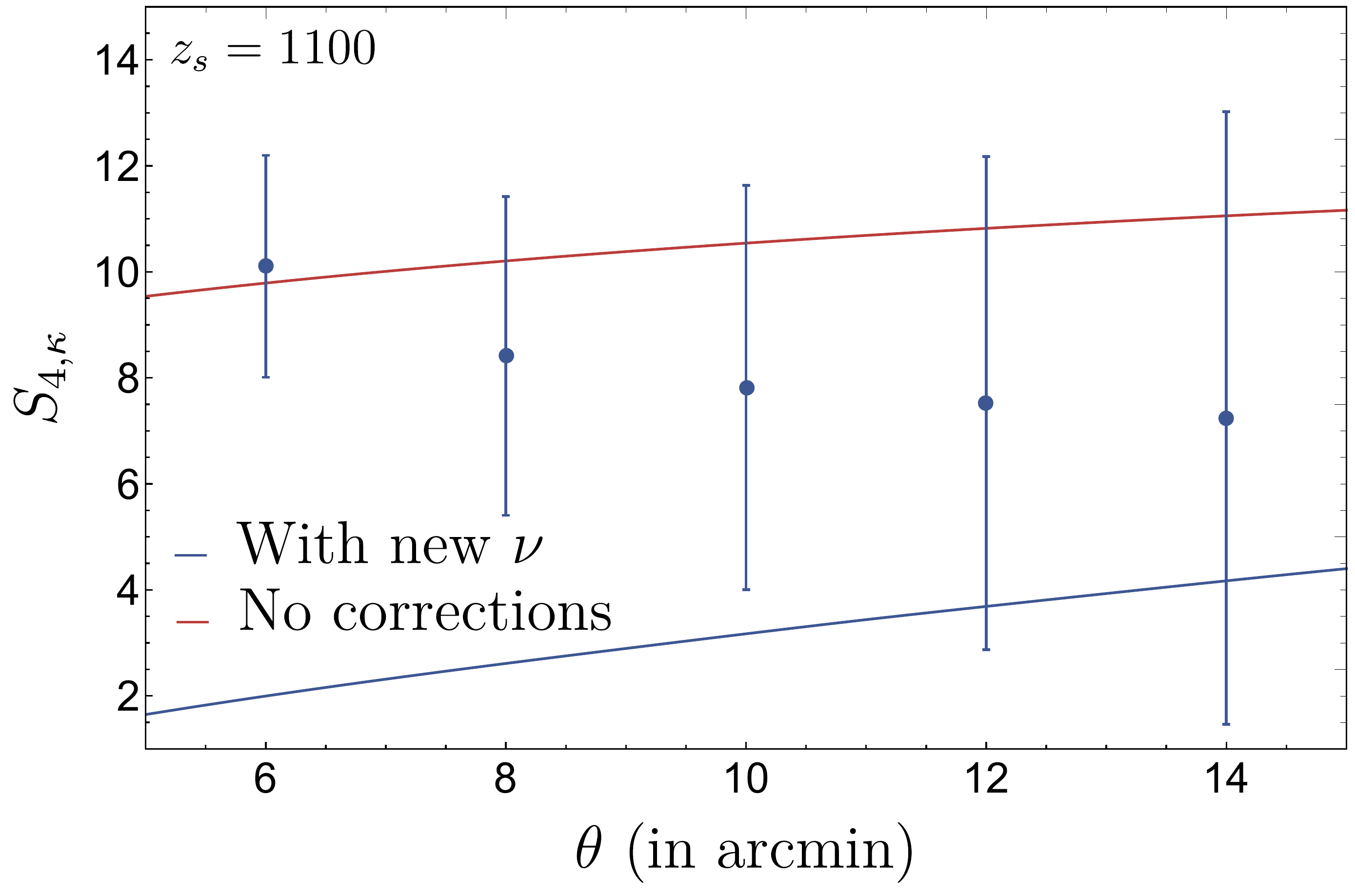}
    \caption{Convergence kurtosis of the CMB with respect to the opening angle of the top-hat smoothing applied. In red is the kurtosis evolution as computed from large deviation theory without taking into account post-Born corrections while in blue these corrections are taken into account via a modification of the collapse parameter in (\ref{collapse}) to match the corrected skewness. The measured points are from the 23 simulated maps (mean of the 23 realisations) described in the main text and error bars represent their standard deviation around the mean.} 
    \label{S4angle}
\end{figure}

To get a more precise idea of the effect of the cumulants on the resulting PDF, 
let us also display separately the CMB lensing skewness (Fig.~\ref{S3angle}) and kurtosis (Fig.~\ref{S4angle}) as a function of the opening angle. 
For the skewness, Fig.~\ref{S3angle} compares measurements in the simulated maps (blue points with error bars given by the dispersion among the mock data) to the prediction from large-deviation theory with (blue solid line) and without (red solid line) post-born corrections. 
For sufficiently large opening angles ($\theta\gtrsim 10$ arcmin), the post-Born corrected prediction is in full agreement with the simulations. 
At smaller scales, deviations start to appear and the measured value of the skewness overshoots the prediction, an effect which might be due to non-linearities and probably call for additional (higher order) corrections.
Note that the {\red previous remark made for the PDF still apply for the skewness itself and we are thus prevented from} testing very precisely (below a few tens of percents) the prediction. {\red When comparing theory and simulated data, let us keep in mind that the error bars for different angular scales are highly correlated, so that overall there is no significant deviation of the measurements from the prediction.} It nonetheless would be very interesting to test in more details the numerical simulation used here although this is not the purpose of this paper.


As for the kurtosis, Fig.~\ref{S4angle} displays the measurements from the simulations against the tree-order prediction without post-Born corrections (red). In addition, we overlay in blue the value of the kurtosis in the formalism developed in this paper that is to say when post-Born corrections {\it to the skewness} are accounted for by modifying the collapse parameter in the large-deviation formalism. Hence,
the red line in Fig.~\ref{S4angle} comes from a physical formalism - large deviation theory witch a meaningful 2D spherical collapse - but without any post-Born corrections while the blue-line is only a consequence of a correction of the spherical collapse parameter only meaningful for the skewness.  
Interestingly, the resulting correction to the kurtosis does not lead to completely unrealistic values (which could have been a real worry and would have made our model unworkable, notably in the tails) and even seems to go in the right direction, lowering the uncorrected value as it should. Given the quite large error bars on the kurtosis, it is difficult to quantify precisely the error we make on this cumulant (do we tend to slightly underestimate it for instance ?). At the level of accuracy of current state-of-the-art CMB lensing simulations, such an effect is completely within the error bars but it would be nonetheless interesting to compare to the tree order kurtosis with post-Born corrections that one could obtain similarly to the skewness but going to next order in the potential. This is left for future works.

\section{Conclusion and discussion}
\label{sec:conclusion}

In this paper, we have computed explicitly within perturbation theory, the leading post-Born corrections to the skewness of the convergence field on mildly non-linear scales. These corrective terms include the so-called lens-lens coupling and geodesic deviation and are shown to be negative, therefore reducing the level of non-Gaussianity.
Even though the correction is small at low redshift as expected, it becomes comparable to the signal at higher redshift, in particular in the context of CMB lensing. Here for apertures about 10 arcminutes, the skewness goes from roughly 2 when post-Born corrections are not accounted for to 1 when they are included. 
We have then added this correction in a prediction for the one-point distribution of the convergence field obtained from large-deviation theory. In order to get a meaningful model both from a mathematical and a physical point of view, the post-Born correction was incorporated at the level of an effective mapping rather than a corrective term in the cumulant generating function which cannot be modelled consistently (this is known as the cumulant problem in mathematics). The model obtained for the PDF of the convergence field was successfully compared to ray tracing simulations, achieving an accuracy below one percent in the five (for negative convergences) to two (for positive convergences) sigmas range around the mean of the distribution for apertures larger than a few arcminutes. This is very encouraging and provides us with a test of  both  CMB lensing simulation -- indeed the simulation used here has its own limitations -- and theory in a regime that has not been investigated in details elsewhere (higher order statistics on mildly non-linear scales).
We have seen that post-Born corrections at the CMB will become relevant in future surveys \citep{Marozzi17} and we emphasised that the lensing one-point PDF is one of the most interesting additional probes to power-spectra, being easy to measure from lensing maps and containing a significant amount of information \citep{JiaCMB}. Indeed, specific forecasts for the Stage-III Advanced Atacama Cosmology Telescope experiment and Stage-IV surveys were performed and showed that even taking into account noise consideration some level of non-Gaussianity can still be measured on the convergence PDF thus making our efforts to model it all the more relevant.

In the future, it would be important to test different numerical schemes and simulations (different ray tracing strategies, resolutions, volumes, etc) and determine how converged/robust is the numerical prediction, specifically in the large convergence tail that may be prone to numerical artefacts as of today. This is however not the purpose of this paper and remains a challenge in the near future given the difficulty for running such simulations.

From a theoretical perspective, it would be of great interest to also try and go beyond the skewness and be able to incorporate post-Born effects in higher order cumulants, for instance the kurtosis and the whole hierarchy of cumulants that govern the tails, although it is not clear whether the latter is within reach of first principle calculations. Such investigation is left for future works.

Finally, let us emphasize that the strategy developed in this paper to incorporate corrective terms into a model for the distribution of a random variable was applied to the particular case of CMB lensing convergence but is very general and could be used also in other contexts.
As an example, one could try and go beyond the tree-order matter density PDF prediction derived in \cite{Bernardeau1992,2014PhRvD..90j3519B,saddle} and add a corrective term to the skewness that would account for its one-loop contribution and therefore might allow us to enlarge the domain of applicability of the model towards possibly smaller scales. This idea will be investigated elsewhere.

\section*{Acknowledgements}
This work is partially supported by the SPHERES grant ANR-18-CE31-0009 of the French {\sl Agence Nationale de la Recherche} and by Fondation MERAC.
AB's work is supported by a fellowship from CNES.
This work has made use of the Horizon Cluster hosted by Institut d'Astrophysique de Paris. We thank Stephane Rouberol for running smoothly this cluster for us, Ken Osato for pointing us to the simulation we used in this paper and Eric Hivon for his help with {\sc healpix}. 
Special thanks go to Rapha\"el Gavazzi for his detailed reading of the manuscript, feedback and support.
We also thank Karim Benabed, Martin Kilbinger,
Aoife Boyle, Oliver Friedrich and Cora Uhlemann 
for fruitful discussions.

\bibliographystyle{mnras}
\bibliography{biblio} 

\appendix
\section{The Sachs equation}
\label{app::sachs}

Here, we rederive the Sachs equation \citep{Sachs61} following some elements from \cite{Bernardeau1997} with some more details for the sake of clarity. We use the signature $(-1,1,1,1)$ and the Einstein's summation convention. 
Note that the derivation proposed here does make use of the usual thin light-beam approximation. It is remarkable that this approximation can be lifted \citep{2017PhRvL.119s1101F,2019PhRvD..99b3525F,2019PhRvD..99b3526F} and although this does not change the range of validity of usual weak-lensing derivations, it does make the link between so-called weak and strong lensing formalisms.

First, let us consider two nearby geodesics $x^{\mu}(\lambda)$ and $x^{\mu}(\lambda)+\xi^{\mu}(\lambda)$ that lie in the past light cone of an observer $O$ and are connected by a vector field $\xi^{\mu}$. $\lambda$ parametrises the geodesics in such a way that we assume the same value at the observer, $\lambda_O = 0$. This allows us to define the wave vector of the photons $k^{\alpha} = {\rm d}x^{\alpha}/{\rm d}\lambda$ which obeys the geodesic equation 
\begin{equation}
    \frac{Dk^{\mu}}{D\lambda} \equiv k^{\alpha}\nabla_{\alpha}k^{\mu} = 0,
    \label{geodesicequation}
\end{equation}
where the covariant derivative along the geodesic $D/D\lambda$ is expressed below in terms of partial derivatives and Christoffel symbols once a
coordinate system is chosen. 
We consider rays with infinitesimal separation and thus the separation vector lies on the null surface $\xi^{\mu}k_{\mu} = 0$ everywhere along the geodesic. In the following we will demonstrate the evolution equation for the separation vector field $\xi^{\mu}$. First let us notice that given an arbitrary coordinate system
\begin{equation}
    \begin{aligned}
    \frac{D\xi^{\mu}}{D\lambda} &= k^{\nu}\nabla_{\nu}\xi^{\mu} ,\\
    &= \frac{{\rm d}}{{\rm d}\lambda}\xi^{\mu} + \Gamma^{\mu}_{\alpha \beta}k^{\alpha}\xi^{\beta} ,\\
    &= k^{\mu}(x+\xi)-k^{\mu}(x) + \Gamma^{\mu}_{\alpha \beta}k^{\alpha}\xi^{\beta}, \\
    \frac{D\xi^{\mu}}{D\lambda} &= \xi^{\nu} \partial_{\nu} k^{\mu}+\Gamma_{\alpha \beta}^{\mu} k^{\alpha} \xi^{\beta} = \xi^{\beta}\nabla_{\beta}k^{\mu}.
    \end{aligned}
\end{equation}
Using this property, let us now obtain an equation of motion for the separation vector
\begin{equation}
    \begin{aligned}
    \frac{D^2\xi^{\mu}}{D\lambda^2} &= k_{\alpha}\nabla^{\alpha}\left(\frac{D\xi^{\mu}}{D\lambda}\right) = k_{\alpha}\nabla^{\alpha}\left[\xi^{\beta}\nabla_{\beta}k^{\mu}\right],\\
    &= k^{\alpha}\left(\nabla_{\alpha}\xi^{\beta}\right)\left(\nabla_{\beta}k^{\mu}\right) + k^{\alpha}\xi^{\beta}\nabla_{\alpha\beta}k^{\mu}, \\
    &= \xi^{\alpha}\left(\nabla_{\alpha}k^{\beta}\right)\left(\nabla_{\beta}k^{\mu}\right) + k^{\alpha}\xi^{\beta}\nabla_{\alpha\beta}k^{\mu}, \\
    &= \xi^{\alpha}\nabla_{\alpha}\left[k^{\beta}\nabla_{\beta}k^{\mu}\right] - \xi^{\alpha}k^{\beta}\nabla_{\alpha \beta}k^{\mu} + k^{\alpha}\xi^{\beta}\nabla_{\alpha\beta}k^{\mu}, \\
    &= \xi^{\alpha}k^{\beta}\left[\nabla_{\beta\alpha}-\nabla_{\alpha \beta}\right]k^{\mu}, \\
    \frac{D^2\xi^{\mu}}{D\lambda^2} &= \xi^{\alpha}k^{\beta}R^{\mu}_{\ \nu\beta\alpha}k^{\nu},
    \label{saachs}
\end{aligned}
\end{equation}
where the last line is obtained using the Ricci identity with $R^{\mu}_{\ \nu\beta\alpha}$ the Riemann tensor. This equation therefore describes the evolution of a light bundle along its geodesic. 
For an observer in $O$ with 4-velocity $v_O^{\mu}$, the angle under which the observer sees the geodesics is given by the components of ${\rm d}\xi^{\mu}/{\rm d}\lambda$ orthogonal to both $k^{\mu}$ and $v_O^{\mu}$ and it is consequently convenient to define two spacelike vectors $n_a^{\mu}$, $a=1,2$, orthogonal to both $k^{\mu}$ and $v_O^{\mu}$ and orthonormal between them that is $g_{\mu \nu} n_{a}^{\nu} n_{b}^{\mu}=\delta_{a b}$ where $g_{\mu\nu}$ is the spacetime metric. Moreover the subspace $\{n_1^{\mu}(\lambda),n_2^{\mu}(\lambda)\}$ is called the \textit{screen} adapted to the observer. This allows us to define a basis at the observer $\left\{n_{1}^{\mu}, n_{2}^{\mu}, k^{\mu}, v_{O}^{\mu}\right\}$ which is extended to all space-time by parallel transporting $n_a^{\mu}$ and $v_O^{\mu}$ along the geodesic. $n_a^{\mu}$ and $v_O^{\mu}$ thus also satisfy geodesic equations such as equation~(\ref{geodesicequation}). As such for all $\lambda$ one can decompose 
\begin{equation}
    \xi^{\mu}=\xi^{0} k^{\mu}+\sum_{a=1,2} \xi^{a} n_{a}^{\mu}+\xi^{u} v_O^{\mu},
\end{equation}
and noticing that the components along $v_O^{\mu}$ is given by $\xi^{\mu}k_{\mu}$ being equal to $0$ what remains is 
\begin{equation}
    \xi^{\mu}=\xi^{0} k^{\mu}+\sum_{a=1,2} \xi^{a} n_{a}^{\mu}.
\end{equation}
Plugging this decomposition into equation~(\ref{saachs}) leads to
\begin{equation}
    \frac{D^2\xi^{\mu}}{D\lambda^2} = R^{\mu}_{\ \nu \alpha \beta}\xi^{a}n_a^{\beta}k^{\alpha}k^{\nu} + R^{\mu}_{\ \nu \alpha \beta}\xi^{o}k^{\beta}k^{\alpha}k^{\nu},
\end{equation}
where the second term is null because $R^{\mu}_{\ \nu \alpha \beta}$ is anti-symmetric in $\alpha\beta$ while $k^\alpha k^\beta$ is obviously symmetric. Thus projecting the equation along the spatial basis $n_a^{\mu}$ remembering that it can enter the derivative because $Dn_a^\mu/D\lambda = 0$ we obtain an evolution equation for the components $\xi^a$
\begin{equation}
    \frac{\mathrm{d}^{2} \xi^{a}}{\mathrm{d} \lambda^{2}}=\mathcal{R}^a_{\ b} \xi^{b},
    \label{evolxi}
\end{equation}
where the symmetric tensor $\mathcal{R}_{ab}$, called optical tidal tensor in what follows, is defined by
\begin{equation}
    \mathcal{R}_{a b} \equiv R_{\mu \nu \rho \sigma} k^{\nu} k^{\rho} n_{a}^{\mu} n_{b}^{\sigma}.
    \label{rab}
\end{equation}
Since equation~(\ref{evolxi}) is linear it is clear that there exists a relation between $\xi^a(\lambda)$, $\xi^a(0) = 0$ and ${\rm d}\xi^a/{\rm d}\lambda(0) = \theta^a_O$ written in the form
\begin{equation}
    \xi^{a}=\hat{\mathcal{D}}_{a b} \theta_{O}^{b}.
    \label{fun}
\end{equation}
Note that $\theta_{O}^{a}$ is the vectorial angle seen by the observer between the two geodesics. $\hat{\mathcal{D}}_{a b}$ is thus the fundamental object we are looking for in lensing, completely describing what we observe, that is the angle (image) we see as opposed to the image on the virtual \textit{screen} moving along the geodesic as defined earlier. We call it the deformation matrix. Finally, plugging equation~(\ref{fun}) into equation~(\ref{evolxi}) leads to a fundamental equation, the Sachs equation, allowing us to compute $\hat{\mathcal{D}}_{a b}$
\begin{equation}
\frac{\mathrm{d}^{2}}{\mathrm{d} \lambda^{2}} \hat{\mathcal{D}}_{a b}=\mathcal{R}_{a}^{\ c} \hat{\mathcal{D}}_{c b}.
\label{Sachs}
\end{equation}
In the case of a non perturbed spacetime, it is clear that the deformation matrix in equation~(\ref{fun}) is the angular distance multiplied by the identity matrix. Since this distance is usually expressed in its comoving version we find convenient to rewrite equation~(\ref{Sachs}) as 
\begin{equation}
    \frac{\mathrm{d}^{2}}{\mathrm{d} \lambda^{2}} a(z) \mathcal{D}_{a b}=a(z)\mathcal{R}_{a}^{\ c}\mathcal{D}_{c b}
\label{Sachs2}
\end{equation}
with $\mathcal{D}_{a b}(0) = 0$ and ${\rm d}\mathcal{D}_{a b}/{\rm d}\lambda(0) = \delta_{ab}$ and where $\delta_{ab}$ is the identity matrix. These conditions express that the geodesics are focused at the observer and that spacetime near the observer is Euclidean.

\section{Post-Born corrections obtained from Fermat's principle}
\label{Sec:Fermat}

Let us start by noticing that at the level of a flat FLRW background, the infinitesimal comoving separation distance in some direction ${\bm \theta}$ at some comoving radial distance $\chi$ ${\bm x}({\bm \theta}, \chi)$ between two geodesics converging at the observer is equal to the comoving radial distance $\chi$ as stated in equation~(\ref{angularcomoving}). This implies that
\begin{equation}
    \frac{{\rm d}^2{\bm x}}{{\rm d}\chi^2} = 0,
    \label{propa}
\end{equation}
which could also be seen as (\ref{evolxi}) written for the background. We now want to include some density perturbations in the propagation equation~(\ref{propa}). If one assumes that the Newtonian potential $\Phi$ is small compared to $c^2$ and that the typical scales over which $\Phi$ changes significantly are much smaller than the curvature scale of the background then according to Fermat's principle each light ray is deflected from its unperturbed path by the transverse gradient of the potential it passes through. Since any physical fiducial ray will be deflected by potential gradients along its way the right-hand side of equation~(\ref{propa}) must thus contain the difference of the perpendicular potential gradients between the two rays to account for the relative deflection of the two rays
\begin{equation}
    \frac{{\rm d}^2{\bm x}}{{\rm d}\chi^2} = -\frac{2}{c^2} \Delta[\nabla_{\perp}\Phi({\bm x}({\bm \theta}, \chi),\chi)].
\end{equation}
This can be solved in a similar fashion as the optical Sachs equation~(\ref{Sachs22}) in the main text and one gets
\begin{multline}
    {\bm x}({\bm \theta}, \chi_s) = \mathcal{D}_0(z_s){\bm \theta} - \frac{2}{c^2}\int_0^{\chi_s} {\rm d}\chi \mathcal{D}_0(z_s-z) \\ \Delta[\nabla_{\perp}\Phi({\bm x}({\bm \theta}, \chi),\chi)].
\end{multline}
Note that contrary to the main text, the transverse gradient operator is here expressed in comoving coordinates. Then going back to the deformation matrix $A_{ij} = \mathcal{D}_{ij}/\mathcal{D}_0$ one gets
\begin{multline}
    A_{ij}({\bm \theta}, \chi_s) = \delta^D_{ij} - \frac{2}{c^2}\int_0^{\chi_s} {\rm d}\chi \frac{\mathcal{D}_0(z_s-z)\mathcal{D}_0(z)}{\mathcal{D}_0(z_s)} \\ \Phi_{,ik}({\bm x}({\bm \theta}, \chi),\chi)A_{kj}({\bm \theta}, \chi).
\end{multline}
The last expression is then explicitly obtained by expanding in parallel $A_{ij} = \delta^D_{ij} + A_{ij}^{(1)}+ ... $ and ${\bm x}({\bm \theta}, \chi_s)$ in powers of $\Phi$. This last expansion corresponds to dropping the Born-approximation since the zeroth order of ${\bm x}({\bm \theta}, \chi_s)$ gives in the previous integral the unperturbed light trajectory while further expanding accounts for the "real" position of lenses. The details of the computation are very similar to what was done in the main text. After some algebra, one eventually finds
\begin{multline}
\label{eq:PBdef}
    A_{ij}^{(2)}({\bm \theta}, \chi_s) = \frac{4}{c^4} \int_0^{\chi_s} {\rm d}\chi \frac{\mathcal{D}_0(z_s-z)\mathcal{D}_0(z)}{\mathcal{D}_0(z_s)} \int_0^\chi {\rm d}\chi' \mathcal{D}_0(z-z') \\ \left[\!\Phi_{,ijl}(\mathcal{D}_0(\chi))\Phi_{,l}(\mathcal{D}_0(\chi'))+\frac{\mathcal{D}_0(\chi')}{\mathcal{D}_0(\chi)}\Phi_{,ik}(\chi)\Phi_{,kj}(\chi')\!\right],
\end{multline}
which is strictly equivalent to what was found in the main text.

\red{}
\section{Sub-optimal approaches to the post-Born PDF}

\subsection{The fast (and dirty) approach}
\label{fast_dirty}

\begin{figure}
    \centering
    \includegraphics[width = \columnwidth]{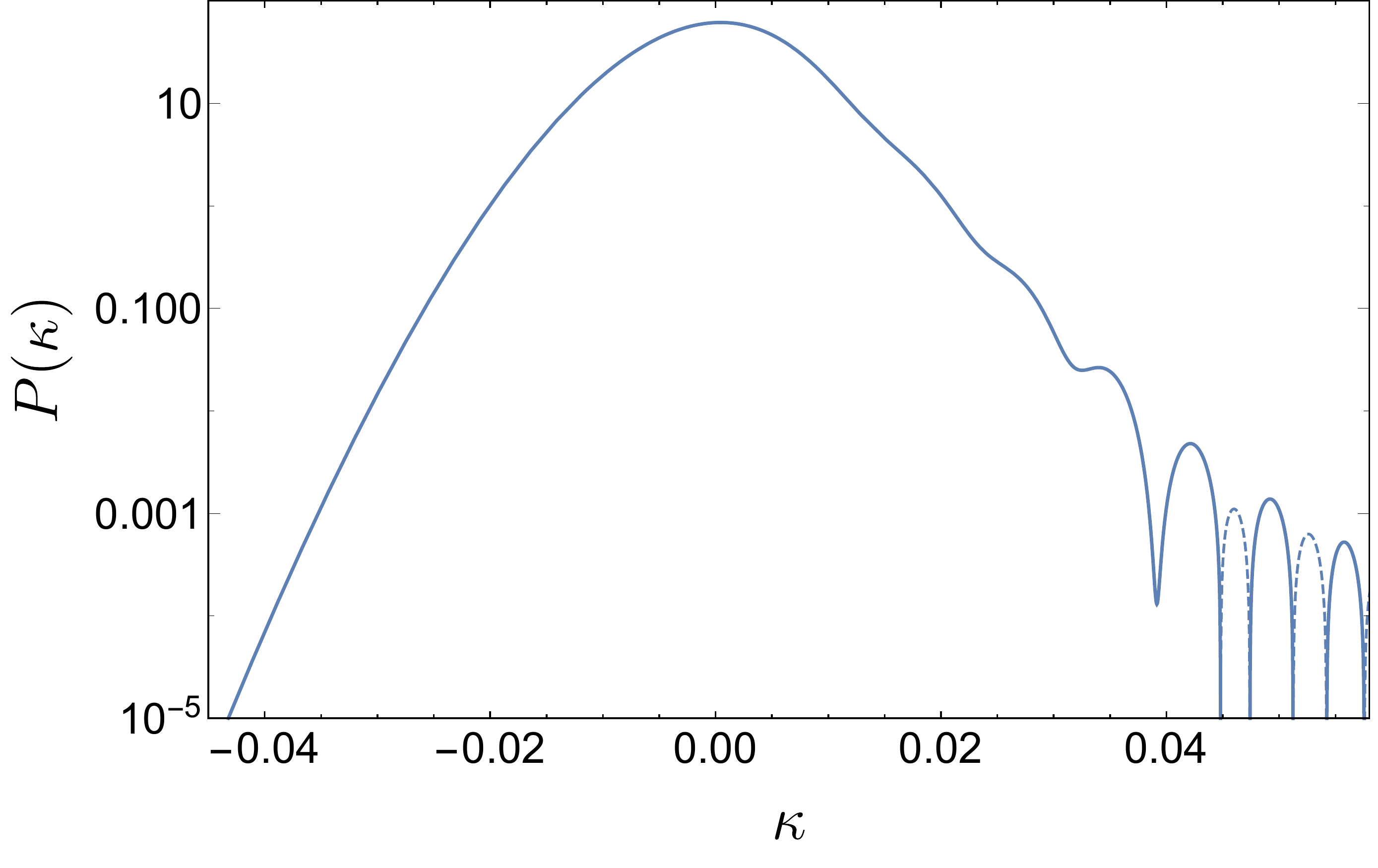}
    \caption{Example of an ad hoc PDF obtained "by hand" after a modification of the skewness using the fast and dirty approach of section~\ref{fast_dirty} that is to say by modifying the cumulant generating function directly. The solid line shows the positive values of the PDF while dashed lines represent the negative values. It is clear that the resulting function displayed on this figure does not possess the elementary mathematical properties of a PDF such as being positive. The highly oscillatory behaviour in the tails is also unphysical unlike what can be typically obtained using large deviation theory.}
    \label{badexample}
\end{figure}
The SCGF whose successive derivatives in zero define the scaled cumulants such as the skewness $S_3$ and the kurtosis $S_4$ can be expressed as 
\begin{equation}
    \varphi_\kappa(y) = \sum_{n=0}^{+\infty} S_n \, \frac{y^n}{n!}.
    \label{defscgf}
\end{equation}
Since this is the key prediction from large deviation theory, and given that we only want to modify $S_3$ for post-Born effects, one attractive option could be to add the post Born correction by hand directly in the SCGF with the following prescription
\begin{equation}
    \varphi_\kappa^{\rm corr}(y) = \sum_{n=0}^{+\infty} S_n \, \frac{y^n}{n!} + S_{3,\kappa}^{\rm corr} \frac{y^3}{6}.
    \label{fastscgf}
\end{equation}
This strategy is found to work well for small corrections to the skewness (as is the case when including primordial non-Gaussianities for instance, see \cite{2019arXiv191206621F}). However this is a shaky construction from a mathematical point of view since there is absolutely no guarantee that the newly defined sequence of cumulants actually corresponds to a random variable, in other words that this new function has all the good properties of a SCGF, in particular to be a PDF once inverse Laplace transformed (positive, normalised, etc). This issue is known as the \textit{cumulant problem}, which states that there is currently no known set of (in)equalities that constrain how any infinite \footnote{A sequence of two numbers, the second being strictly positive, always corresponds to a Gaussian, otherwise a cumulant generating function must contain an infinite number of non zero cumulants, the variance always being a strictly positive value.} sequence of real numbers can represent the sequence of cumulants of a random variable ($11^{\rm th}$ problem in \cite{Rota1998}). 

This can also be hinted using standard results of perturbation theory by showing that the cumulants are all inter-dependent and thus one cannot simply modify one without touching consistently the others. 
Indeed it was showed
\citep{Bernardeau1995,BernardeauReview} that the reduced cumulants $S_n$ as given by the spherical collapse model (i.e at tree order) are all expressed as a function of the so-called vertices $\nu_i$ of order $i=n$ {\it and less}. These vertices appear in the spherical collapse model as the Taylor coefficients of the collapse dynamics seen as a mapping between the non-linear (late-time) density and the initial density contrast.
For example the skewness and kurtosis at tree order read\footnote{Note that this is strictly valid only in the case of an Einstein-de-Sitter background but found to be an excellent approximation for standard FLRW cosmology.}
\begin{equation}
\begin{aligned}
S_3^{2D}(R) &= 3 \nu_2 + \frac{3}{2} \, \frac{{\rm d}\log(\sigma^2(R))}{{\rm d}\log(R)}, \\
S_4^{2D}(R) &= 4\nu_3 + 12\nu_2^2 + (14\nu_2-2) \, \frac{3}{2} \, \frac{{\rm d}\log(\sigma^2(R))}{{\rm d}\log(R)}  \\ & \quad + \frac{21}{4} \, \left(\frac{{\rm d}\log(\sigma^2(R))}{{\rm d}\log(R)}\right)^2 + \frac{{\rm d^2}\log(\sigma^2(R))}{{\rm d}\log(R)^2} ,
\end{aligned}
\end{equation}
in a 2D slice of radius $R$ and where the values of the vertices $\nu_n$ are tabulated for the exact 2D spherical collapse ($\nu_2=12/7$, $\nu_3=29/7$, etc).
This illustrates the need for a joint modification of {\it all} cumulants simultaneously in order to get a consistent modelling of the full one-point statistics. If one changes the value of the skewness (to correct for the post-Born effects for instance), one could change accordingly the value of the other cumulants so as to keep the interpretation in terms of an underlying mapping (or vertex structure).
Let us finally show a concrete example of the issue with Fig.~\ref{badexample} which displays the result one would obtain when trying to alter the skewness in an arbitrary PDF\footnote{Here, we choose the PDF of the convergence for a source redshift $z_s=1$ and aperture $\theta=10$ arcmin for which we substracted 70 to the original skewness of 77.} using this approach. It is clear that this function, although each successive integral yields the correct values of the cumulants (this is actually used to check that the inverse Laplace transform was performed successfully), cannot be considered a correct PDF since displaying negative values and a highly oscillating behaviour in the tails unless what large deviation theory typically produces.
In order to avoid this issue, we abandon this strategy and propose another approach in the next section which allows us to get meaningful predictions from a mathematical and physical point of view. 

\subsection{Post-Born effect on each lens}

\begin{figure}
    \centering
    \includegraphics[width = \columnwidth]{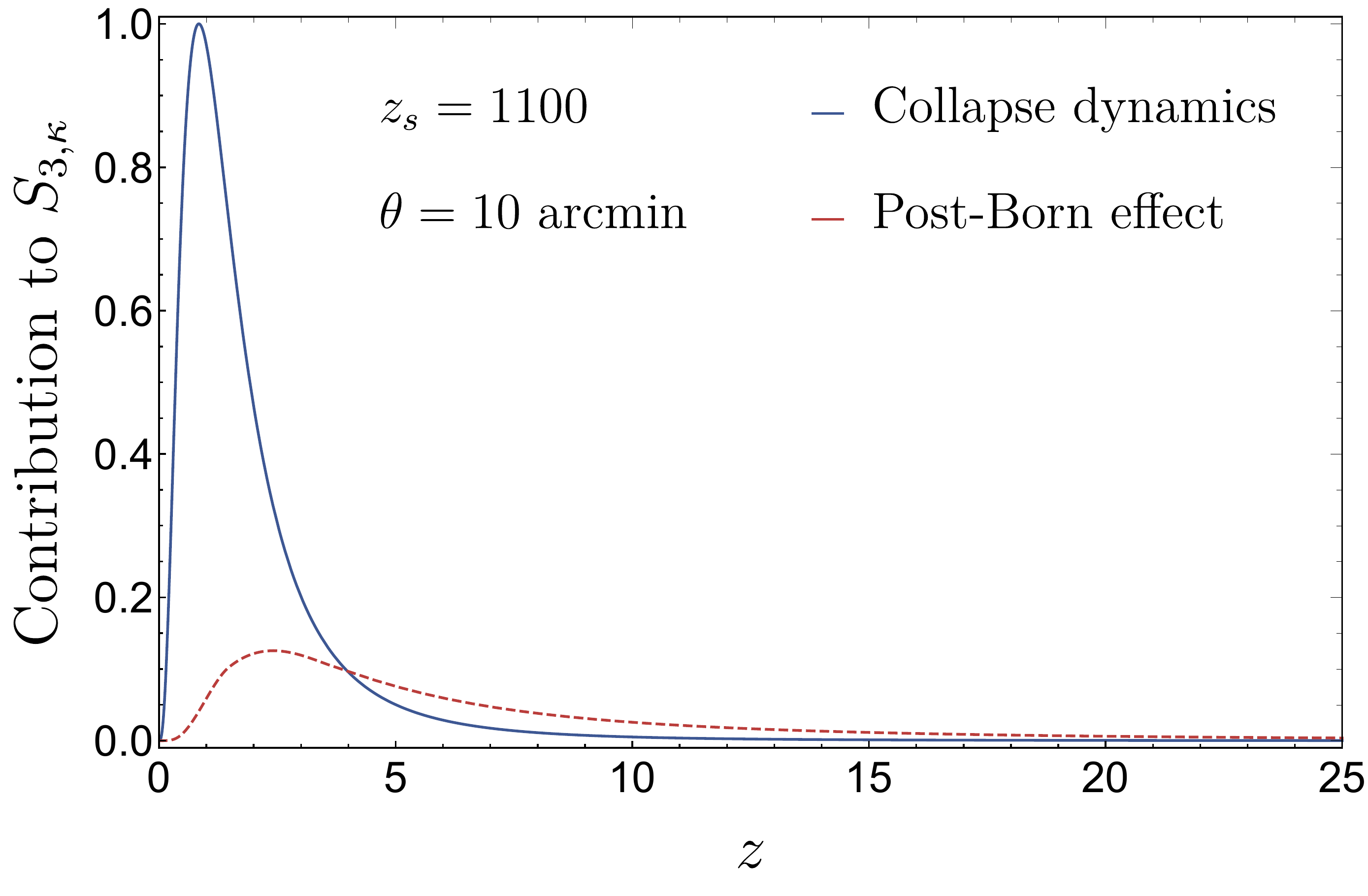}
    \caption{Absolute contributions to the convergence skewness coming from each slice and from the 2D spherical collapse dynamics (blue) and the post-Born corrections (red). Dashed line represent negative values. Here the contributions are shown for a source redshift of $z_s = 1100$ and an opening angle of $\theta = 10$ arcmin.}
    \label{contributions}
\end{figure}

What actually ensures that our PDFs are indeed PDFs is the formal construction from large deviation theory where all the cumulants are obtained from a "most probable" mapping between the initial conditions of the field and the field at late time. The approach followed in \cite{Barthelemy19} consists in the description of an implicit mapping between the initial convergence field and the final field where the mapping is built from the dynamics of the underlying density field, however it is effectively simply a formal mapping for the convergence. Therefore, since taking into account post-Born corrections must in a sense be equivalent to finding a more appropriate mapping, we can change the mapping of the underlying density field to ensure that it effectively encompasses those corrections for the convergence. 

Another method than used in the main text could consist in modifying, for each density slice that makes the convergence, the value of the 2D spherical collapse parameter $\nu$ in equation~(\ref{collapse}) so as to reproduce the tree-order corrected convergence skewness instead of the uncorrected one. In physical terms, changing the mapping along the line-of-sight would account for the fact that close lenses act as a deformation of background lenses and thus alter the collapse the observer sees. Indeed, the correction displayed in equation~(\ref{finalk3corr}) can be seen as a line-of-sight correction to the third cumulant of each density slice and thus a correction of the collapse parameter $\nu$. Thus if one sees the correction as the integral of some function $F(z,z_s)$,
\begin{equation}
    \left\langle\kappa^3_{\rm corr}\right\rangle_\theta = \int_0^{z_s} \frac{{\rm d}z}{H(z)} F(z,z_s),
    \label{rewritek3}
\end{equation}
and uses equation~(14) of \cite{Barthelemy19} to get the uncorrected tree-order skewness, one can write the corrected convergence as
\begin{equation}
    \left\langle\kappa^3\right\rangle_\theta = \int_0^{z_s} \frac{{\rm d}z}{H(z)} \omega(z,z_s)^3 \left(\left\langle \delta^3 \right\rangle + \frac{F(z,z_s)}{\omega(z,z_s)^3} \right),
\end{equation}
where $\left(\left\langle \delta^3 \right\rangle + F(z,z_s)/\omega(z,z_s)^3 \right)$ will lead to the new skewness value of the slice and hence the corrected collapse parameter. Note that since $F(z,z_s)$ is itself an integral from the observer to the slice it indeed states how foreground lenses affect each density slice. However for the mapping to make sense in each slice along the line-of-sight, that is keeping the functional of the spherical collapse, it is mandatory that the collapse dynamics still dominates over the post-Born corrections, in other words that the contribution to the convergence skewness coming from the gravitational collapse in each slice is more important than the one coming from the integrand of equation~(\ref{rewritek3}). Indeed, it would not make sense to keep the functional describing collapse dynamics as written in equation~(\ref{collapse}) if it is not anymore the dominant effect.

We plot in Fig.~\ref{contributions} the contribution to the skewness of each slice depending on the redshift of the slice for a source redshift $z_s=1100$ and an opening angle $\theta = 10$ arcmin. This plot shows that the contribution from the post-Born effects, although overall subdominant, dominates the dynamics of the slices for a large fraction of the line-of-sight. This leads to negative values of $\nu$ in many slices which totally changes the shape of a the mapping. As such, although this approach and the one displayed in the main text would give rather similar PDFs -- same mean, variance and skewness --, keeping the (modified) spherical collapse mapping for all the slices would not make any sense and would lead, from a physical point of view, to an overall bad construction.

\label{lastpage}
\end{document}